\documentclass[sigconf]{acmart}
\usepackage{tikz-cd}
\usetikzlibrary{decorations.pathmorphing,shapes}

\usepackage{mathtools}

\theoremstyle{definition}
\newtheorem{definition}{Definition}[section]
\newtheorem{theorem}{Theorem}

\newtheorem{example}{Example}
\newtheorem{msfoltheory}{MSFOL Theory}[theorem]

\author{Hoang Nguyen}
\email{hoang.nguyen@inf.ethz.ch}
\affiliation{%
\institution{%
ETH Zurich}
\city{Zurich}
  \country{Switzerland}
}
\thanks{Part of this work was done at IMDEA Software Institute, Madrid, Spain.}

\author{Manuel Clavel}
\email{manuel.clavel@eiu.edu.vn}
\affiliation{%
\institution{%
Eastern International University}
\city{Binh Duong}
  \country{Vietnam}
}

\keywords{Object Constraint Language, SQL, Correctness}

\title{Proving correctness for SQL implementations of OCL constraints}

\begin{abstract}
In the context of the model-driven development of data-centric applications, 
OCL constraints play a major role in adding precision 
to the source models (e.g., data models and security models). 
Several code-generators have been proposed 
to bridge the gap between source models with OCL constraints and their corresponding database implementations. 
However, the database queries produced by
these code-generators are significantly less efficient---from the point
of view of execution-time performance--- than the  implementations manually written by  database experts. 
In this paper, we propose
a different approach to bridge the gap between models with
OCL constraints and their corresponding database 
implementations. In particular, 
we introduce a model-based methodology for proving 
the correctness of manually written
SQL implementations of OCL constraints. 
This methodology is
based on a novel mapping from 
a significant subset of the SQL language
into many-sorted first-order logic.  
Moreover, by leveraging on an already existing mapping
from the OCL language into many-sorted first-order logic,
we can use SMT solvers to automatically prove the correctness
of SQL implementations of OCL constraints. 
To illustrate and show the applicability of our approach,
we include in the paper a number of non-trivial
examples. Finally, we  report on the  status 
of a suite of tools supporting our approach.
\end{abstract}

\begin{document}
\maketitle

\section{Introduction}
In the context of software development, model-driven engineering (MDE)~\cite{DBLP:series/synthesis/2017Brambilla} aspires
to develop software systems by using models as the driving-force. Models are
artefacts defining the different aspects and views of the intended software system.
Ideally, the gap between the source models and the real software systems is
covered by appropriate code-generators.

The Unified Modelling Language (UML)~\cite{UML17} is the facto standard modeling language for MDE. Originally, it was conceived as a graphical language: models
were defined using diagrammatic notation. However, it  soon became clear 
that UML diagrams were not expressive enough to define certain aspects of
the intended software systems, 
and the Object Constraint Language (OCL)~\cite{OCL14}
was added to the UML standard. 

OCL is a textual language, with a formal
semantics. It can be used to specify in a precise, unambiguous way complex constraints and queries over models. In the context of model-driven development of
 data-centric applications, OCL has been  used to specify both data models' invariants
and security models' authorisation constraints~\cite{DiosDBC14}.

A number of mappings from OCL to other languages (e.g.,~\cite{DBLP:conf/vecos/BennamaB14, DBLP:conf/models/Bergmann14}) have been
proposed in the past, each with its own goals and limitations. 
In the context of the model-driven development of data-centric applications, in order
to bridge the gap between models with
OCL constraints and their corresponding database 
implementations, \cite{DemuthH99,HeidenreichWD08,
EgeaDC10,EgeaD19,BaoC19}
introduce different mappings from OCL to SQL. 
Unfortunately, as reported in~\cite{ClavelB19,BaoC19}, 
the SQL queries produced by
these mappings are significantly less 
efficient ---from the point of view
 of execution-time performance--- than the corresponding implementations  written by  SQL experts.

In this paper, we follow
a different approach for bridging the gap between models with
OCL constraints and their corresponding database 
implementations. 
We assume that, in practice, OCL constraints are
implemented by SQL experts. Then, to bridge the 
aforementioned gap,
we propose a model-based methodology for proving 
the correctness of  these implementations.
However, 
proving  correctness for SQL implementations of OCL constraints poses a number of  non-trivial challenges. In particular, although
both languages can be considered as query languages, the
``resources'' that they provide for specifying  queries are
of different nature. Suffice to say that there is really 
nothing like OCL iterators  in the standard 
SQL language.\footnote{Interestingly, for the case of 
mapping OCL iterator expressions,~\cite{EgeaDC10,EgeaD19} 
propose using
cursors and loops within
stored procedures. But this is certainly not
 the natural way of implementing queries
 in SQL and, not-surprisingly, it comes with a significant 
 penalty in terms of execution-time efficiency.}
Moreover, while the {\tt Boolean} type in OCL has four values
---namely, {\tt true}, {\tt false}, {\tt null}, and {\tt invalid}---,
in SQL it has only three values ---namely,
{\tt TRUE},  {\tt FALSE} and {\tt NULL}.
Furthermore, the null-value behaves differently 
in OCL and in SQL. As an example, 
the expression ${\tt null} = {\tt null}$ in OCL 
evaluates to {\tt true} (much as in
object-oriented programming languages, like Java), 
while executing in SQL
the statement {\tt SELECT NULL = NULL} returns 
{\tt NULL}.

Our methodology is
based on a novel mapping, called SQL2\-MSFOL, from 
a significant subset of the SQL language
into many-sorted first-order logic (MSFOL), which takes into
account the aforementioned challenges.
Then, by leveraging on an existing mapping
from OCL  into MSFOL~\cite{DaniaC16},
we can use Satisfiability Modulo Theories (SMT) solvers~\cite{BarrettCDHJKRT11, MouraB08} 
to automatically prove the correctness
of the SQL implementations of OCL constraints.

\paragraph{Organisation}
In Section~\ref{bird-eye-view:sec} we
provide a bird's-eye-view of our methodology, indicating the mappings that we use and the roles that they play.
Then, in Section~\ref{datamodels_to_msfol:sec}
we recall the  
mappings that we borrow from the literature and
use in our methodology.
Next, in Section~\ref{sql_to_msfol:sec} we
introduce the key component of our methodology:
namely, a novel mapping from SQL to 
many-sorted first-order logic (MSFOL).
Afterwards, in Section~\ref{examples:sec}
we discuss a number of non-trivial
examples of correctness proofs
that illustrate our 
methodology, 
and in Section~\ref{tools:sec} 
we report on the  status 
of a suite of tools supporting our methodology.
Finally, in Sections~\ref{related-work:sec}
and~\ref{future-work:sec}
we discuss related work and future work.
For the sake of readability, we have moved
to the appendices the detailed definitions of the 
different mappings. 

\section{A bird's-eye view}
\label{bird-eye-view:sec}
\begin{figure}
\begin{eqnarray*}
\begin{tikzcd}[column sep=small]
\mathcal{D}  \arrow[rrrrd, bend left, "{\rm o2f}()"]
\arrow[dd, "{\rm o2s}()"']  \arrow[r, dashleftarrow]
& {\it expr}   \arrow[dd, rightsquigarrow] \arrow[rr, "{\rm o2f}()"] 
& & {\rm o2f}({\it expr})  \arrow[rd, dashrightarrow]  \arrow[dd, "?" description, Leftrightarrow]
&  \\ 
& & & & 
{\rm o2f}(\mathcal{D}) \cup {\rm s2f}({\rm o2s}
(\mathcal{D}))
\\
{\rm o2s}(\mathcal{D})  \arrow[rrrru, bend right, "{\rm s2f}()"']
\arrow[r, dashleftarrow] 
& {\it sel} \arrow[rr, "{\rm s2f}()"']  
& & {\rm s2f}({\it sel})  \arrow[ru, dashrightarrow]
&  \\ 
\end{tikzcd}
\end{eqnarray*}
\caption{A schematic depiction of our proposal}
\label{diagram}
\end{figure}
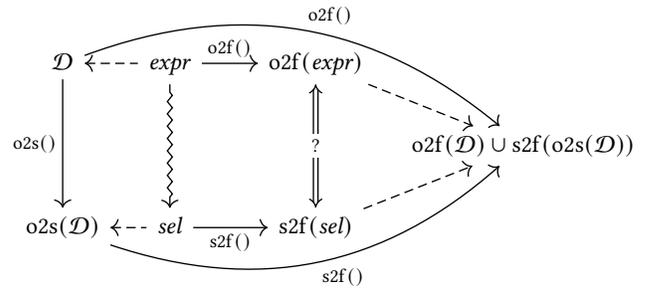

In Figure~\ref{diagram} we
depict schematically our methodology for
proving the correctness of SQL implementations of
OCL constraints.
We briefly indicate in this section  
the mappings that we use and the 
roles that they play in our methodology.
A detailed account of each mapping is given
in the following sections.

First, our methodology leverages on a previous mapping,
called OCL2MSFOL~\cite{DaniaC16}, from OCL to
many-sorted first-order logic.
We denote the mapping OCL2MSFOL as ${\rm o2f}()$.
Let 
$\mathcal{D}$ be a data model and let ${\it expr}$ be an OCL Boolean
expression in the context of $\mathcal{D}$.
In a nutshell, the mapping ${\rm o2f}()$  generates
\begin{itemize}
\item a  MSFOL theory ${\rm o2f}(\mathcal{D})$ such that there is a 
correspondence between the instances of 
the data model 
$\mathcal{D}$ and the models of the theory
${\rm o2f}(\mathcal{D})$;
\item a  MSFOL formula ${\rm o2f}({\it expr})$ such that,
for any instance $\mathcal{O}$ of $\mathcal{D}$, the expression ${\it expr}$ evaluates to {\tt true}
in the instance $\mathcal{O}$ if and only if the formula ${\rm o2f}({\it expr})$ holds in the 
model of  the theory ${\rm o2f}(\mathcal{D})$ that
corresponds to the instance $\mathcal{O}$.
\end{itemize}

Secondly, our methodology  uses the mapping 
OCL2\-PSQL from OCL to SQL~\cite{BaoC19}.
We denote  the part of the mapping
OCL2PSQL related to data models as ${\rm o2s}()$.
Let  $\mathcal{D}$ be a data model.
In a nutshell, the mapping ${\rm o2s}()$  generates
\begin{itemize}
\item a  SQL schema ${\rm o2s}(\mathcal{D})$ 
such that there is a 
correspondence between the instances of 
the data model 
$\mathcal{D}$ and the database instances of the schema
${\rm o2s}(\mathcal{D})$.
\end{itemize}

Finally, our methodology introduces 
a new mapping, called SQL2MSFOL, from
SQL to many-sorted first-order logic.
We denote SQL2MSFOL as ${\rm s2f}()$.
Let  $\mathcal{D}$ be a data model.
Let ${\rm o2s}(\mathcal{D})$ be the SQL schema
corresponding to $\mathcal{D}$ 
and let ${\it sel}$ be a SQL select-statement in the context of
${\rm o2s}(\mathcal{D})$.
In a nutshell, the mapping ${\rm s2f}()$  generates
\begin{itemize}
\item a  MSFOL theory ${\rm s2f}({\rm o2s}(\mathcal{D}))$
such that there is a 
correspondence between the database instances of
the schema 
${\rm o2s}(\mathcal{D})$ and the models of the theory
${\rm o2f}(\mathcal{D})\cup {\rm s2f}({\rm o2s}(\mathcal{D}))$;
\item a  MSFOL formula ${\rm s2f}({\it sel})$ such that,
for any database instance $\mathcal{Y}$ of ${\rm o2s}(\mathcal{D})$,
the result of executing the statement ${\it sel}$ in $\mathcal{Y}$ is {\tt TRUE}
 if and only if the formula ${\rm s2f}({\it sel})$ holds in the 
model of the theory  
${\rm o2f}(\mathcal{D})\cup {\rm s2f}({\rm o2s}(\mathcal{D}))$ 
that corresponds to the database instance $\mathcal{Y}$.
\end{itemize}

Let  $\mathcal{D}$ be a data model
and let ${\it expr}$ be an OCL Boolean
expression in the context of $\mathcal{D}$.
 Let ${\it sel}$ be a SQL select-statement in the context of
${\rm o2s}(\mathcal{D})$.
In our  methodology, we consider that ${\it sel}$ correctly implements
the Boolean expression ${\it expr}$ if and only if the following holds:
\begin{eqnarray*}
{\rm o2f}(\mathcal{D}) \cup {\rm s2f}({\rm o2s}
(\mathcal{D}))
\models {\rm o2f}({\it expr})\Longleftrightarrow {\rm s2f}({\it sel}).
\end{eqnarray*}

\section{Mapping OCL to MSFOL}
\label{datamodels_to_msfol:sec}
The Object Constraint Language (OCL)~\cite{OCL14} is a language for specifying
constraints and queries using a textual notation.
Every OCL expression is written in the context of a model
(called the contextual model). 
OCL is strongly typed. Expressions
either have a primitive type, a class type, a tuple type, or a
collection type. OCL provides standard operators on primitive types,
tuples, and collections. 
OCL also provides a
dot-operator to access the value of  
an attribute of an object, or the  objects linked with an object at the end
of an association. 
OCL provides operators to iterate over collections, 
such as \texttt{forAll}, 
\texttt{exists}, 
\texttt{select}, 
\texttt{reject}, and
\texttt{collect}.
Collections can be sets, bags, 
ordered sets and sequences, 
and  can be parametrised by any type, including
other collection types.
Finally, to represent \emph{undefinedness},
OCL provides two constants, namely, \texttt{null} and
\texttt{invalid}. 
Intuitively, \texttt{null} represents
an unknown or an undefined value, whereas \texttt{invalid} represents
an error or an exception.

\subsubsection*{Data models and object models}
We assume that the contextual models of OCL expressions
are basic data models consisting of  classes, 
with attributes, and  associations. The interested reader can
find our formal definitions of data models and  object models
(or instances of data models)
in Appendix~\ref{dm-om:app}.

OCL expressions are evaluated in object models. 
Let $\mathcal{D}$ be a data model. 
We denote by ${\rm Inst}(\mathcal{D})$ the set of
instances of $\mathcal{D}$.
Now, let $\mathcal{O}\in {\rm Inst}(\mathcal{D})$ 
be an instance of $\mathcal{D}$. Let ${\it expr}$ be an OCL expression.
Then, we denote by ${\rm Eval}(\mathcal{O}, {\it expr})$ the evaluation of ${\it expr}$ in $\mathcal{O}$ according to the 
semantics of OCL.

\subsubsection*{Assignments}
The standard semantics of OCL can be naturally extended
to take into account \emph{assignments} as follows.

A variable $x$ in an OCL expression ${\it expr}$ is \emph{free} if it is not bound by any iterator in ${\it expr}$. We denote by 
${\rm Free}({\it expr})$ the set  of free variables  occurring in ${\it expr}$.
%

Let $\mathcal{D}$ be a data model. Let $\mathcal{O}\in {\rm Inst}(\mathcal{D})$ 
be an instance of $\mathcal{D}$. Let $x$ be an OCL variable. Let $v$ be
an OCL value. 
Then, $x \mapsto v$ is a
\emph{valid assignment} 
if and only 
if $x$ and $v$ are of the same OCL type.
Let $\vec{X}$ be a set of OCL variables.
We denote by 
${\rm Asg}(\vec{X}, \mathcal{O)}$ the set of valid \emph{assignments} of 
values in $\mathcal{O}$ to variables in $\vec{X}$.

Let $\mathcal{D}$ be a data model. Let $\mathcal{O}\in {\rm Inst}(\mathcal{D})$ 
be an instance of $\mathcal{D}$. 
Let ${\it expr}$ be an OCL expression,
with free variables $\vec{X}$. 
Let $\sigma\in {\rm Asg}(\vec{X}, \mathcal{O})$ be a valid assignment for
$\vec{X}$. 
Then, we denote by ${\rm Eval}(\mathcal{O}, \sigma, {\it expr})$ the evaluation of ${\it expr}$ in $\mathcal{O}$ according to the 
semantics of OCL, where  variables in $\vec{X}$ 
are evaluated following the assignment $\sigma$.

\subsubsection*{Mapping OCL to MSFOL}
\cite{DaniaC16} defines a mapping, called OCL2MS\-FOL,
 from OCL Boolean
expressions  to many-sorted first-order (MSFOL) formulas. 
More precisely, OCL2MSFOL consists
of a mapping
${\rm o2f}_{\rm data}()$  from data models  to  MSFOL theories,
and  four mappings, namely, ${\rm o2f}_{\rm true}()$,
${\rm o2f}_{\rm false}()$, ${\rm o2f}_{\rm null}()$, and ${\rm o2f}_{\rm inval}()$, formalising, respectively, when an OCL Boolean expression evaluates to ${\tt true}$, ${\tt false}$,  ${\tt null}$,
or ${\tt invalid}$. 

The interested reader can find a description of these
mappings in Appendix~\ref{ocl2msfol:app}.
Nevertheless, there are two  properties of the aforementioned mappings that we should remark here.

First, let $\mathcal{D}$ be a data model. Then,
there is a one-to-one correspondence 
between the instances 
of the data model $\mathcal{D}$ and the models 
of the corresponding MSFOL theory 
${\rm o2f}_{\rm data}(\mathcal{D})$.

Secondly, the four mappings ${\rm o2f}_{\rm true}()$,
${\rm o2f}_{\rm false}()$, ${\rm o2f}_{\rm null}()$, and ${\rm o2f}_{\rm inval}()$ are defined recursively over the
structure of OCL expressions.
When the subexpression is
a non-Boolean type, an auxiliary mapping 
${\rm o2f}_{\rm eval}()$ is called.
As expected, the 
mapping ${\rm o2f}_{\rm eval}()$ builds upon the mapping 
${\rm o2f}_{\rm data}()$.
However, there are three  classes of
non-Boolean expressions that ${\rm o2f}_{\rm eval}()$
treats especially.
First, the class of expressions that define sets
(e.g., {\tt allInstances}-expressions, 
and {\tt select}, {\tt reject}, and {\tt collect}-expressions).
Each expression
${\it expr}$ in this class is mapped by 
${\rm o2f}_{\rm eval}()$ to a new predicate.
This predicate formalises the set defined by the
expression ${\it expr}$ and its definition is generated 
as part of the mapping ${\rm o2f}_{\rm eval}()$.
Second, the class of expressions that distinguish an element from a set (e.g., {\tt any}, {\tt max}, and {\tt min}-expressions). 
Each expression {\it expr} in this class is mapped by 
${\rm o2f}_{\rm eval}()$ to a new function.
This function  represents the element referred to by ${\it expr}$
and its definition is generated 
as part of the mapping ${\rm o2f}_{\rm eval}()$.
Finally,  the class of literal expressions. 
For each literal  
in ${\it expr}$, the mapping  ${\rm o2f}_{\rm eval}()$ generates
the axioms stating that this literal is different from ${\tt null}$ and
${\tt invalid}$.\footnote{This is needed because 
OCL2MSFOL maps  the
${\tt null}$ and ${\tt invalid}$ for each predefined type to
distinguished  constants of the 
corresponding predefined types in MSFOL.}

We denote by 
${\rm o2f}_{\rm def}({\it expr})$ 
the union of the set of axioms generated
by ${\rm o2f}_{\rm eval}()$, when considering
each non-Boolean (sub)expres\-sion of ${\it expr}$.

Finally, let $\mathcal{D}$ be a data model.
Let $\mathcal{O}$ be an object model
 of $\mathcal{D}$ and let ${\rm intr}(\mathcal{O})$
 be the corresponding model in  ${\rm o2f}_{\rm data}(\mathcal{D})$.
 Let ${\it expr}$ be a ground (i.e., no free variables)
 OCL Boolean expression.  Then,~\cite{DaniaC16} claims
 that the following holds:
%
\begin{eqnarray*}
\label{correctness-map-true}
\lefteqn{{\rm Eval}(\mathcal{O}, {\it expr}) = {\tt true}}\\
& &  \Longleftrightarrow
{\rm intr}(\mathcal{O})\models 
({\rm o2f}_{{\rm def}}({\it expr})
\Rightarrow
{\rm o2f}_{{\rm true}}({\it expr}))
\end{eqnarray*}

This result can be naturally extended to take into account assignments. Let $\mathcal{D}$ be a data model and let $\mathcal{O}$ be an object model
 of $\mathcal{D}$. Let ${\it expr}$ be a
 OCL Boolean expression, with free variables ${\vec X}$.
 Then, for any assignment $\sigma \in {\rm Asg}(\vec{X}, \mathcal{O})$, 
%
\begin{eqnarray*}
\label{correctness-map-true-asg}
\lefteqn{{\rm Eval}(\mathcal{O}, \sigma, {\it expr}) = {\tt true}}\\
& & \Longleftrightarrow
{\rm intr}(\mathcal{O}, \sigma)\models 
({\rm o2f}_{{\rm def}}({\it expr})
\Rightarrow
{\rm o2f}_{{\rm true}}({\it expr}))
\end{eqnarray*}
%
\noindent where ${\rm intr}(\mathcal{O}, \sigma)$ denotes the interpretation 
${\rm intr}(\mathcal{O})$ extended with the assignment $\sigma$.


The key limitations of the mapping OCL2MSFOL
come from the fact
that expressions defining collections are mapped to predicates. 
Although these new predicates
are defined so as to capture the property that distinguishes
the elements belonging to the given collection, this is not
sufficient for reasoning about this collection's cardinality, or
about the multiplicity or the ordering of its elements. As a consequence, 
the mapping OCL2MSFOL cannot support, in general, {\tt size}-expressions or expressions of collection types different from set types.

\section{Mapping SQL to MSFOL}
\label{sql_to_msfol:sec}
The Structure Query Language (SQL) is a special-purpose programming language designed for managing data in relational database management systems (RDBMS). Its scope includes data insert, query, update and delete, and schema creation and modification.

\subsubsection*{Our mapping's context}
Our methodology for proving the correctness of 
SQL implementations of OCL constraints is based on
the mapping  SQL\-2MSFOL explained below.
Notice that, in particular,  our mapping assumes that the SQL implementations
of the OCL constraints are select-statements in 
the context of the SQL schema generated from the
contextual  model  of the OCL constraints.

More specifically, we borrow from~\cite{BaoC19}
the mapping from  data models to
SQL schemata, which we denote as  ${\rm o2s}()$,
 and the corresponding mapping
 from instance of data models
to instances of SQL databases, which we denote as
${\rm o2s}_{\rm inst}()$.
In a nutshell, the mapping ${\rm o2s}()$ maps
classes to tables, attributes to columns, and many-to-many associations to tables with appropriate foreign-keys.
The mapping ${\rm o2s}_{\rm inst}()$ maps objects and
links accordingly.
In what follows, 
let $\mathcal{D}$ be a data model. 
Then, for any
class $c$ in $\mathcal{D}$ we denote by 
$\ulcorner c\urcorner$ the table in ${\rm o2s}(\mathcal{D})$
corresponding to the class $c$.
Similarly, for any
association ${\it as}$ in $\mathcal{D}$ we denote by 
$\ulcorner {\it as}\urcorner$ the table in ${\rm o2s}(\mathcal{D})$
corresponding to the association ${\it as}$.
The interested reader can find the description of
the mappings ${\rm o2s}()$ and ${\rm o2s}_{\rm inst}()$ in Appendix~\ref{ocl2psql:app}.
Nevertheless, there is a property of these mappings 
that we should remark here.
Let $\mathcal{D}$ be a data model. Then, there is a one-to-one correspondence between the instances of the data model 
$\mathcal{D}$ and the
database instances of the schema ${\rm o2s}(\mathcal{D})$.

\subsubsection*{Assignments}
Let $\mathcal{S}$ be a database schema. 
We denote by ${\rm Inst}(\mathcal{S})$ the 
set of database instances of $\mathcal{S}$.
%
Let $\mathcal{Y}\in {\rm Inst}(\mathcal{S})$ be a database
instance of $\mathcal{S}$.
Let $x$ be a SQL variable.
Let $v$ be
a SQL value (in the context of $\mathcal{Y}$). Then, $x \mapsto v$ is
a \emph{valid assignment} 
if and only 
if $x$ and $v$ are of the same SQL type.
Let $\vec{X}$ be a set of SQL variables.
We denote by 
${\rm Asg}(\vec{X}, \mathcal{Y)}$ the set of valid \emph{assignments} of 
values in $\mathcal{Y}$ to variables in $\vec{X}$.

SQL statements are executed on database instances. 
Let $\mathcal{S}$ be a database schema. 
Let $\mathcal{Y}\in {\rm Inst}(\mathcal{S})$ 
be a database instance of $\mathcal{S}$. Let ${\it sel}$ be a 
SQL select-statement
Let $\varsigma \in {\rm Asg}(\vec{X}, \mathcal{Y})$ be a valid
assignment for $\vec{X}$. 
Then, we denote by ${\rm Exec}(\mathcal{Y}, \varsigma, {\it sel})$ 
the result of executing ${\it sel}$ in $\mathcal{Y}$ according to the 
semantics of SQL, where the variables in $\vec{X}$ are substituted according to the assignment $\varsigma$.

Let $\mathcal{D}$ be a data model. Let
$\mathcal{O}\in {\rm Inst}(\mathcal{D})$.
Let $\vec{X}$ be a set of variables,
and let $\sigma\in {\rm Asg}(\vec{X}, \mathcal{O})$ be
a valid assignment for $\vec{X}$ (in the context of $\mathcal{O}$).
Then, we denote by ${\rm o2s}_{\rm inst}({\sigma})\in {\rm Asg}(\vec{X}, {\rm o2s}_{\rm inst}(\mathcal{O}))$ the
assignment of values in ${\rm o2s}_{\rm inst}(\mathcal{O})$ to the variables in
$\vec{X}$ that correspond, according to the mapping
${\rm o2s}_{\rm inst}()$, 
to the values assigned 
by $\sigma$ to the variables in $\vec{X}$.

\subsubsection*{Our notion of correctness}
We are interested in proving the correctness of
SQL implementations of  
OCL constraints, i.e., of
OCL expressions of type {\tt Boolean}.
As mentioned before, the type
{\tt Boolean} in OCL has four values: ${\tt true}$, 
${\tt false}$, ${\tt null}$, and ${\tt invalid}$.
Our notion of correctness
for SQL implementations of OCL constraints is  only
 concerned  with the case when
the OCL constraints evaluate to ${\tt true}$.
This is arguably the  most interesting case, in practice, when
using OCL constraints.

More specifically, let $\mathcal{D}$ be a data model.
Let ${\it expr}$ be an OCL Boolean expression, with free variables $\vec{X}$. Let ${\it sel}$ be a SQL select-statement
containing exactly
one expression ${\it selitem}$ in its list of selected items.
We consider that ${\it sel}$  
correctly implements ${\it expr}$ if and only if,
for every instance $\mathcal{O} \in {\rm Inst}(\mathcal{D})$ and every 
assignment $\sigma \in {\rm Asg}(\vec{X}, \mathcal{O})$, the following holds:
${\rm Eval}(\mathcal{O}, \sigma, {\it expr})$ evaluates to 
${\tt true}$ if and only if
${\rm Exec}({\rm o2}_{\rm inst}(\mathcal{O}), 
{\rm o2}_{\rm inst}(\sigma), {\it sel})$ returns ${\tt TRUE}$.

\subsubsection*{Our mapping in a nutshell}
Let $\mathcal{D}$ be a data model.
Let $\mathcal{O} \in  {\rm Inst}(\mathcal{D})$ be an instance
of  $\mathcal{D}$.
Let ${\it sel}$ be a SQL select-statement
containing exactly
one expression ${\it selitem}$ in its list of selected items.
In a nutshell, the mapping SQL2MSFOL defines the following:

\begin{itemize}
\item For each class $c$ in $\mathcal{D}$,  our mapping defines
a predicate ${\rm index}_{\ulcorner c\urcorner}()$ that
specifies the \emph{indices} (of the rows)
of the table $\ulcorner c\urcorner$ in $\mathcal{O}$.
Then, for each attribute ${\it att}$ of $c$  
our mapping defines a function ${\rm val}_{\ulcorner c\urcorner}({\it att}, x)$ 
that specifies  the  \emph{value} of the column ${\it att}$ 
in the row indexed by $x$
in the table $\ulcorner c\urcorner$.

\item For each association ${\it as}$ in $\mathcal{D}$,  our mapping defines
a predicate ${\rm index}_{\ulcorner {\it as}\urcorner}()$ that
specifies the \emph{indices} (of the rows)
of the table $\ulcorner {\it as} \urcorner$ in $\mathcal{O}$.
Then, for each association-end ${\it ase}$ of ${\it as}$ 
our mapping defines a function ${\rm val}_{\ulcorner {\it as}\urcorner}({\it ase}, x)$ 
that specifies  the  \emph{value} of the column ${\it ase}$ 
in the row indexed by $x$
in the table $\ulcorner{\it as}\urcorner$.

\item For each (sub)select ${\it sel'}$ in  ${\it sel}$, our mapping defines
a predicate ${\rm index}_{\it sel'}()$ that
specifies the \emph{indices} (of the rows)
of the table ${\rm Exec}({\rm o2s}_{\rm inst}(\mathcal{O}), 
{\it sel'})$.
\item For each (sub)expression ${\it expr}$ in each 
(sub)select ${\it sel'}$ in ${\it sel}$, 
our mapping defines a function ${\rm val}_{\it sel'}({\it expr}, x)$ 
that specifies  the  \emph{value} of the expression ${\it expr}$ 
in the row indexed by $x$
in the table 
${\rm Exec}({\rm o2s}_{\rm inst}(\mathcal{O}), {\it sel'})$.
Notice that the value of the expression ${\it expr}$
can be a Boolean value, which in SQL is either
{\tt TRUE}, {\tt FALSE}, or {\tt NULL}. To represent
the SQL Boolean values, our mapping
generates an enumerated type with the literals
$\underline{{\tt TRUE}}$, $\underline{{\tt FALSE}}$, 
and $\underline{{\tt NULL}}$. 

\end{itemize}
  
We denote by 
${\rm index}_{\rm def}(\mathcal{D})$ 
the union of the set of axioms specifying
the predicates ${\rm index}_{\ulcorner c\urcorner}()$
and ${\rm index}_{\ulcorner {\it as}\urcorner}()$,
as well as the functions
${\rm val}_{\ulcorner c\urcorner}({\it att})$
and  ${\rm val}_{\ulcorner {\it as}\urcorner}({\it ase})$
 for every  class $c$,
 attribute ${\it att}$, association ${\it as}$,
 and association-end ${\it ase}$
in $\mathcal{D}$.

We denote by 
${\rm index}_{\rm def}({\it sel})$ 
the union of the set of axioms specifying
the predicate ${\rm index}_{\it sel'}()$,
for every  (sub)select-statement ${\it sel'}$  in  ${\it sel}$,
Similarly, we denote by 
${\rm val}_{\rm def}({\it sel})$ 
the union of the set of axioms specifying
the function ${\tt val}_{\it sel'}({\it expr})$,
for every (sub)expression ${\it expr}$ of every 
(sub)select-statement ${\it sel'}$  in  ${\it sel}$.

The interested reader can find the formal definition of our mapping  in Appendix~\ref{sql2msfol:app}.
The task of formally proving that SQL2MSFOL
is  correct, i.e., that  it correctly captures the semantics of SQL, 
is beyond the scope of this paper.

\subsubsection*{Re-formalising our notion of correctness using
our mapping}
We can now use our mapping SQL2MSFOL, along with
the mapping  OCL2MSFOL~\cite{DaniaC16}, to
re-formalise our notion of correctness 
for SQL implementations of
OCL constraints as follows.

Let $\mathcal{D}$ be a data model.
Let ${\it expr}$ be an OCL Boolean expression, with free variables $\vec{X}$. Let ${\it sel}$ be a SQL select-statement
containing exactly
one expression ${\it selitem}$ in its list of selected items.
We consider  that   ${\it sel}$ is a correct implementation
of ${\it expr}$ if and only if the  MSFOL 
theories~C1,~C2 and~C3 below, with the variables $\vec{X}$
added as (uninterpreted) constants of 
the appropriate type, are \emph{unsatisfiable}.
We precede each theory by its intended meaning. 

\begin{msfoltheory} 
\label{theory:C1}
There exists at least one instance $\mathcal{O}$
of $\mathcal{D}$ such that: 
the query ${\it sel}$ returns zero or more than one
row when executed in  
${\rm o2s}_{\rm inst}(\mathcal{O})$. Formally,
\begin{eqnarray*}
\lefteqn{{\rm o2f}_{\rm data}(\mathcal{D})
}\\
& & \cup\  {\rm index}_{\rm def}(\mathcal{D}) 
\cup {\rm index}_{\rm def}({\it sel})  
\cup {\rm val}_{\rm def}({\it sel})\\
& & \cup\ \{\neg(\exists(x).({\rm index}_{\it sel}(x)
\wedge \forall(y).(y \not= x \Rightarrow 
\neg({\rm index}_{\it sel}(y)))))\}.
\end{eqnarray*}
\end{msfoltheory}

\begin{msfoltheory} \label{theory:C2}
There exists at least one instance
of $\mathcal{D}$ such that: (i) the expression
${\it expr}$ evaluates to {\tt true} in $\mathcal{O}$,
and
(ii) there exists at least one row in the table obtained
 when executing ${\it sel}$ in ${\rm o2s}(\mathcal{O})$
 for which  ${\it selitem}$ does not contain
the value {\tt TRUE}.
\begin{eqnarray*}
\lefteqn{{\rm o2f}_{\rm data}(\mathcal{D}) 
\cup {\rm o2f}_{\rm def}({\it expr})}\\
& &\cup\  {\rm index}_{\rm def}(\mathcal{D}) 
\cup {\rm index}_{\rm def}({\it sel})  
\cup {\rm val}_{\rm def}({\it sel})\\
& & \cup\ \{{\rm o2f}_{\rm true}({\it expr})\}\\
& & \cup\ \{\neg(\forall(x).({\rm index}_{\it sel}(x)
\Rightarrow {\rm val}_{\it sel}({\it selitem}, x)
= \underline{{\tt TRUE}}))\}.
\end{eqnarray*}
\end{msfoltheory}

\begin{msfoltheory} \label{theory:C3}
There exists at least one instance
of $\mathcal{D}$ such that: (i)
 ${\it selitem}$ contains
the value {\tt TRUE} in
all the rows of the table obtained
 when executing ${\it sel}$ in ${\rm o2s}(\mathcal{O})$,
 and (ii) the expression
${\it expr}$ does not evaluate to {\tt true} in $\mathcal{O}$.
\begin{eqnarray*}
\lefteqn{{\rm o2f}_{\rm data}(\mathcal{D})
\cup {\rm o2f}_{\rm def}({\it expr})}\\
& &\cup\  {\rm index}_{\rm def}(\mathcal{D}) 
\cup {\rm index}_{\rm def}({\it sel})  
\cup {\rm val}_{\rm def}({\it sel})\\
& & \cup\ \{\forall(x).({\rm index}_{\it sel}(x)
\Rightarrow {\rm val}_{\it sel}({\it selitem}, x)
= \underline{{\tt TRUE}})\}\\
& & \cup\ 
 \{\neg({\rm o2f}_{\rm true}({\it expr})\}.\\
\end{eqnarray*}
\end{msfoltheory}



\section{Examples}
\label{examples:sec}

\label{examples:sec}

To illustrate and show the applicability of our approach,
we include in this section a number of non-trivial
examples of proving correctness of SQL implementations of
OCL constraints. 

\begin{figure}
\centering
\includegraphics[width=0.45\textwidth]{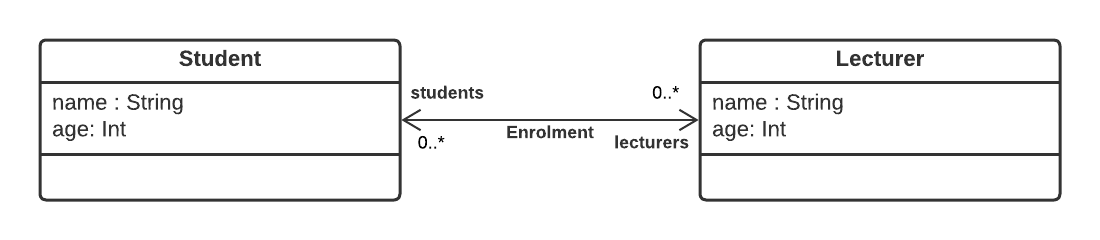}
 \caption{The data model ${\tt University}$.}
 \label{university:dm}
 \end{figure}

Consider the
data model {\tt University}  in Figure~\ref{university:dm}.
It contains two classes, 
{\tt Student} and {\tt Lecturer}, and an association,
{\tt Enrolment}. 
{\tt Student} and {\tt Lecturer}  represent, respectively, the students and the lecturers of the university.
{\tt Student} and {\tt Lecturer} have attributes {\tt name} and {\tt age}. 
{\tt Enrolment} represents the relationship between  a student (at the association-end {\tt students}) and a lecturer 
(at the association-end  {\tt lecturers})  when the student is enrolled in a course taught by the lecturer.

The interested reader can find 
the correctness proofs for the examples below
at~\cite{submission-artifacts}.
Notice that the file names in this examples repository follow
the pattern {\tt exm}$i${\tt -C}$j$, where the integer $i$ 
denotes  the number of
the example (from 1 to 7)  under consideration,
and the integer $j$ denotes  the number of
the  theory (from 1 to 3)
in Definition~C1,~C2 and~C3 whose satisfiability is checked.
In other words, 
for each example below,
and for each theory in Definition~C1,~C2,~C3,
 the interested reader
can find at~\cite{submission-artifacts} a file specifying, 
using SMT-LIB2 syntax~\cite{BarST-SMT-10},
the corresponding satisfiability problem,
and can check whether this problem  is satisfiable or not  
by inputting the file into
any SMT solver that supports the SMT-LIB2 language,  
like CVC4~\cite{BarrettCDHJKRT11} 
or Z3~\cite{MouraB08}.


\subsection*{Example\#1}
Consider the OCL expression:
\begin{tabbing}
{\tt >} \texttt{true.}
\end{tabbing}

Using our methodology we can prove that
the  SQL statement below correctly
implements the above  OCL expression:
\begin{tabbing}
{\tt >} {\tt SELECT TRUE;}
\end{tabbing}

In particular, for the corresponding correctness proof,
we use the files
{\tt exm1-C1},
{\tt exm1-C2}, and {\tt exm1-C3}
at~\cite{submission-artifacts}.

\subsection*{Example~\#2}

Consider the OCL expression:
\begin{tabbing}
{\tt >} {\tt caller.students}$\rightarrow${\tt isEmpty().}
\end{tabbing}
where  {\tt caller} is a variable of
type {\tt Lecturer}.

Using our methodology we can prove that
the  SQL statement below correctly
implements the above  OCL expression:

\begin{tabbing}
{\tt >} {\tt SELECT NOT EXISTS}\\
{\tt >} \qquad{\tt (SELECT students FROM Enrolment}\\
{\tt >} \qquad{\tt WHERE lecturers = caller);}
\end{tabbing}

In particular, for the corresponding correctness proof,
we use the files
{\tt exm2-C1},
{\tt exm2-C2}, and {\tt exm2-C3}
at~\cite{submission-artifacts}.

\subsection*{Example~\#3}
Consider the OCL expression:

\begin{tabbing}
{\tt >} ${\tt self.age}\geq 18$.
\end{tabbing}
where  {\tt self} is a variable of
type {\tt Student}.

Using our methodology we can prove that
the  SQL statement below correctly
implements the above  OCL expression:

\begin{tabbing}
{\tt >} {\tt SELECT age >= 18 FROM Student}\\
{\tt >} \quad{\tt WHERE Student\_id = self;}
\end{tabbing}

In particular, for the corresponding correctness proof,
we use the files
{\tt exm3-C1},
{\tt exm3-C2}, and {\tt exm3-C3}
at~\cite{submission-artifacts}.

\subsection*{Example~\#4}
Consider the OCL expression:

\begin{tabbing}
{\tt >}  {\tt Student.allInstances()}$\rightarrow$\\
{\tt >} \quad {\tt forAll(s}$\mid$ {\tt s.lecturers}$\rightarrow${\tt forAll(l}
$\mid$
{\tt s.age} $<$ {\tt l.age))}.
\end{tabbing}

Using our methodology we can prove that
the  SQL statement below correctly
implements the above  OCL expression:
\begin{tabbing}
{\tt >} {\tt SELECT NOT EXISTS}\\
{\tt >} \quad{\tt (SELECT 1 FROM}\\
{\tt >} \qquad{\tt (SELECT s.age, e.lecturers}\\
{\tt >} \qquad{\tt \ FROM Student s JOIN Enrolment e}\\
{\tt >} \qquad{\tt \ ON e.students = s.Student\_id) AS TEMP}\\ 
{\tt >} \quad{\tt JOIN Lecturer l}\\
{\tt >} \quad{\tt WHERE TEMP.age >= l.age}\\
{\tt >} \quad{\tt AND l.Lecturer\_id = TEMP.lecturers);}
\end{tabbing}

In particular, for the corresponding correctness proof,
we use the files
{\tt exm4-C1},
{\tt exm4-C2}, and {\tt exm4-C3}
at~\cite{submission-artifacts}.

\subsection*{Example~\#5}
Consider the OCL expression:

\begin{tabbing}
{\tt >} {\tt self.name = user}.
\end{tabbing}
where  {\tt self} is a variable of
type {\tt Student}, and
{\tt user} is a variable of type {\tt String}.

Suppose  that the variable {\tt user} is always assigned a 
{\tt String} value different from {\tt null}.
Then, using our methodology we can prove that
the  SQL statement below correctly
implements the above  OCL expression:

\begin{tabbing}
{\tt >} {\tt SELECT (SELECT name FROM Student}\\
{\tt >} \quad{\tt WHERE Student\_id = self) = user;}
\end{tabbing}

In particular, for the corresponding correctness proof,
we use the files
{\tt exm5-C1},
{\tt exm5-C2}, and {\tt exm5-C3}
at~\cite{submission-artifacts}.
Notice that we have added the assumption about the
variable {\tt user} being assigned a 
{\tt String} value different from {\tt null} 
 to each of the satisfiability problems
in  {\tt exm5-C1},
{\tt exm5-C2}, and {\tt exm5-C3}.
Otherwise, given the SQL semantics for
{\tt null}-equality, we can not prove that
this SQL statement  correctly implements 
the OCL expression above, since
there are scenarios where the OCL expression
 will evaluate to {\tt true} but the
SQL statement will not return ${\tt TRUE}$.
For example,
suppose that the variable
{\tt user} is assigned {\tt null}.
Then, for an scenario in which the {\tt name}
of the lecturer {\tt self} is also {\tt null}, 
the OCL expression
will evaluate to {\tt true} while the SQL statement will
return {\tt NULL}.

\subsection*{Example~\#6}
Consider the same OCL expression as in Example~\#5.

\begin{tabbing}
{\tt >} {\tt self.name = user}.
\end{tabbing}
where  {\tt self} is a variable of
type {\tt Student}, and
{\tt user} is a variable of type {\tt String}.

Suppose, as in Example~\#5,  that the variable
{\tt user} is always assigned a 
{\tt String} value different from {\tt null}.
Consider now the SQL statement:

\begin{tabbing}
{\tt >} {\tt SELECT name = user FROM Student}\\
{\tt >}\quad{\tt WHERE Student\_id = self;}
\end{tabbing}

\noindent Notice that this  statement is different 
(but equivalent) to
the statement in Example~\#5.
As expected, using our methodology we can prove that
this SQL statement  correctly
implements the above OCL expression.
In particular, for the corresponding correctness proof,
we use the files
{\tt exm6-C1},
{\tt exm6-C2}, and {\tt exm6-C3}
at~\cite{submission-artifacts}.
Notice that, as in Example~\#5 and for the same reasons,
we have added the assumption about the
variable {\tt user} being assigned a 
{\tt String} value different from {\tt null} 
 to each of the satisfiability problems
in  {\tt exm6-C1},
{\tt exm6-C2}, and {\tt exm6-C3}.

\subsection*{Example~\#7}
Consider again 
the same OCL expression as in Example~\#5 and 
Example~\#6.

\begin{tabbing}
{\tt >} {\tt self.name = user}.
\end{tabbing}
where  {\tt self} is a variable of
type {\tt Student}, and
{\tt user} is a variable of type {\tt String}.

However, in this example, we do not assume that 
the variable
{\tt user} is always assigned a 
{\tt String} value different from {\tt null}.
Consider now the SQL statement:

\begin{tabbing}
{\tt >} {\tt SELECT CASE WHEN name IS NULL}\\
{\tt >} \quad{\tt THEN user IS NULL}\\
{\tt >} \quad{\tt ELSE CASE WHEN user IS NULL}\\
{\tt >} \qquad{\tt THEN FALSE}\\
{\tt >} \qquad{\tt ELSE name = user END}\\
{\tt >} \quad{\tt END}\\
{\tt >} \quad{\tt FROM Student WHERE Student\_id = self;}
\end{tabbing}

Notice that in this statement the {\tt case}-expressions 
take care of
the ``problematic'' scenarios. As expected, 
using our methodology we can prove that
this SQL statement correctly
implements the above  OCL expression
In particular, for the corresponding correctness proof,
we use the files
{\tt exm7-C1},
{\tt exm7-C2}, and {\tt exm7-C3}
at~\cite{submission-artifacts}.

\section{Tool support}
\label{tools:sec}
We have implemented the following tools to automate the transformation process (i.e., text-to-model, model-to-text)
supporting our methodology.

\paragraph{OCL2MSFOL}\cite{tool:OCL2MSFOL} is a 
 Java application that implements the mapping OCL2MSFOL
from OCL to many-sorted first-order logic (MSFOL)~\cite{DaniaC16}.
Given a data model $\mathcal{D}$ (in {\tt JSON}-format) and an OCL constraint ${\it expr}$ (in textual format), OCL2MSFOL generates the MSFOL theory 
${\rm o2f}(\mathcal{D})$ and the 
MSFOL formula ${\rm o2f}({\it expr})$ 
(in {\tt SMT-LIB2} syntax). 
\paragraph{SQL2MSFOL}\cite{tool:SQL2MSFOL} is a Java implementation of our mapping SQL2\-MSFOL
from SQL statements to many-sorted first-order logic (MSFOL). 
Given a data model $\mathcal{D}$ (in {\tt JSON}-format) and an SQL select-statement ${\it sel}$ (in textual format), 
SQL2MSFOL automatically generates the
 MSFOL theory ${\rm s2f}(\mathcal{D})$ and the MSFOL formula ${\rm s2f}({\it sel})$ (in {\tt SMT-LIB2} syntax).
\paragraph{OCLSQLProver}\cite{tool:integration} 
is a Python 
application that integrates the tools OCL2MSFOL
and SQL2MSFOL. In a nutshell, OCLSQLProver
takes
a data model $\mathcal{D}$, an OCL constraint
${\it expr}$, a set (possibly empty) of OCL assumptions,
and a 
SQL select-statement ${\it sel}$, and it
automatically generates
the satisfiability problems that, according to our methodology,
need to be checked for proving that the SQL select-statement
${\it sel}$ correctly implements the OCL constraint
${\it expr}$. 
Then, OCLSQLProver will call
an SMT solver of choice (i.e., CVC4~\cite{BarrettCDHJKRT11}, Z3~\cite{MouraB08}) 
to check the generated satisfiability problems.
If, for every generated
satisfiability problem, the result is {\tt UNSAT},
 we can conclude the SQL select-statement
${\it sel}$ correctly implements the OCL constraint
${\it expr}$. If, for some of
the generated satisfiability problems, the result is {\tt SAT},
we can conclude that  the
SQL select-statement
${\it sel}$ does not correctly implements the OCL constraint
${\it expr}$. Finally, if, for some of
the generated satisfiability problems, the result is {\tt UNKNOWN}, then we cannot conclude 
whether the SQL select-statement
${\it sel}$ correctly implements or not the OCL constraint ${\it expr}$.

\section{Related work}
\label{related-work:sec}

To the best of our knowledge, there have been no previous 
proposals for proving correctness of SQL implementations of OCL constraints. On the other hand, a number of different
 mappings have been proposed in the past to generate SQL implementations from 
OCL expressions~\cite{DemuthH99,HeidenreichWD08,
EgeaDC10,EgeaD19,BaoC19}.
The interested reader can find in~\cite{ClavelB19,BaoC19}
a detailed discussion about the goals and limitations of
each of these mappings. No formal proof of the correctness of
the aforementioned mappings have been published yet.

\looseness=-1
Our methodology for proving  correctness of
SQL implementations of OCL constraints crucially leverages on the mapping
OCL2\-MSFOL~\cite{DaniaC16} from OCL to many-sorted first-order logic. 
The interested reader can  find in~\cite{DaniaC16} a summary of previously
proposed mappings from OCL to other formalisms. 
Among these mappings, 
USE~\cite{GogollaBR07,GogollaBC13} and HOL-OCL~\cite{BruckerW08} 
are the ones more
closely related to OCL2MSFOL. 
For reasoning
about UML/OCL models, 
USE uses SAT-based constraint solvers
and HOL-OCL uses the interactive theorem prover Isabelle~\cite{DBLP:books/sp/NipkowPW02, tool:Isabelle},
 while OCL2MSFOL uses SMT solvers.
On the other hand, HOL-OCL supports the full
OCL language, while OCL2MSFOL only supports
a significant subset of the language.

Finally, the key component of our methodology 
is our  mapping
SQL\-2MSFOL from SQL to many-sorted first-order logic.
Although designed for different purposes, it would be  
interesting to compare the mapping SQL\-2MSFOL with the formal semantics
(for a basic fragment) of SQL introduced in~\cite{GuagliardoL17}. 
The interested reader can  find in~\cite{GuagliardoL17} a detailed
discussion about previous attempts of providing 
formal semantics to SQL.

\section{Conclusions and future work}
\label{future-work:sec}
In the context of  model-driven development of 
data-centric applications, 
OCL constraints can play a major role in adding precision 
to the source models. In particular, OCL has been
successfully used  to specify both data models' invariants
and security models' authorization constraints~\cite{DiosDBC14}.

A number of
code-generators have been proposed in the past
to bridge the gap between source models with OCL constraints and their corresponding SQL implementations~\cite{DemuthH99,HeidenreichWD08,
EgeaDC10,EgeaD19,BaoC19}.
Unfortunately, the database queries produced by
these code-generators are significantly less efficient---from the point
of view of execution-time performance--- than 
the  corresponding implementations written by  SQL experts~\cite{ClavelB19,BaoC19}.

To bridge the gap between source models 
with OCL constraints and their corresponding SQL implementations, we have proposed in this paper
a different approach. Namely, instead of 
generating the SQL implementations from the OCL
constraints using  code-generators ---and  relying for
their correctness on the correctness of the code-generators
themselves---, we 
propose a methodology
for proving the correctness of the SQL implementations 
themselves ---letting the SQL experts to decide
the most efficient way of implementing the OCL constraints.

Our methodology is
based on a novel mapping, called SQL\-2MSFOL, from 
a significant subset of the SQL language
into many-sorted first-order logic.  
Moreover, by leveraging on the mapping OCL\-2MSFOL~\cite{DaniaC16}
from the OCL language into many-sorted first-order logic,
we can use SMT solvers (e.g., CVC4~\cite{BarrettCDHJKRT11} 
or Z3~\cite{MouraB08}) to automatically prove the 
correctness of SQL implementations of OCL constraints. 
Moreover, we have included in this paper a number of non-trivial
examples that show the applicability of our methodology,
and we have briefly reported on the  status 
of a suite of tools supporting it.

As for future work, the first natural direction is to extend our mapping
SQL2MSFOL to include more features of the SQL language, especially
aggregation and grouping.  
Nevertheless, our main future work will consist in formally proving the correctness of
our mapping SQL2MSFOL, at least for the basic fragment of SQL covered by~\cite{GuagliardoL17}.
For the proof of correctness of SQL2MSFOL, we intend to use  interactive theorem provers like Isabelle~\cite{DBLP:books/sp/NipkowPW02, tool:Isabelle} or Coq~\cite{DBLP:series/txtcs/BertotC04, tool:Coq}.

\section*{Acknowledgment}

Hoang Nguyen is supported by the Swiss National Science Foundation grant ``Model-driven Security \& Privacy'' (204796). 

\bibliography{mybib}
\bibliographystyle{ACM-Reference-Format}

\appendix
\section{Data models and object models}
\label{dm-om:app}
We define data models as follows:

\begin{definition}
Let $\mathcal{T} = \{{\tt Integer}, {\tt String}\}$ be the set of predefined types.
A \emph{data model} $\mathcal{D}$ is a tuple
$\langle \mathit{C}, \mathit{AT}, \mathit{AS}\rangle$, where:
\begin{itemize}

\item $C$ is a set of \emph{classes} $c$.

\item $\mathit{AT}$ is
  a set of \emph{attributes}  ${\it att}$, ${\it att} = \langle \mathit{ati}$, $c$, $t\rangle$,
  where ${\it ati}$ is the attribute's identifier,
  $c$ is the class of the  attribute,
  and $t$ is the type of the values of the attribute, with $t\in \mathcal{T}$ or $t\in C$.

\item $\mathit{AS}$ is
  a set of \emph{associations}   ${\it as}$,
  ${\it as} = \langle {\it asi}, \mathit{ase}_{\rm l}$, $c_{\rm l}$,
  $\mathit{ase}_{\rm r}$, $c_{\rm r}\rangle$,
  where ${\it asi}$ is the association's identifier,
  ${\it ase}_{\rm l}$ and ${\it ase}_{\rm r}$ are the association's ends,
  and ${c}_{\rm l}$ and ${c}_{\rm r}$ are the classes of the objects at the
  corresponding association's ends.
\end{itemize}

\end{definition}


Then, we define instances of data models as follows:
\begin{definition}
Let ${\mathcal D}$ = $\langle \mathit{C}, \mathit{AT}, \mathit{AS}\rangle$ 
be a data model. 
An \emph{object model} $\mathcal{O}$ of $\mathcal{D}$ (also called an \emph{instance} of 
$\mathcal{D}$)
is a tuple $\langle \mathit{OC}$, $\mathit{OAT}$, $\mathit{OAS}\rangle$
where:
\begin{itemize}
\item $\mathit{OC}$ is set of 
  objects $o$,   $o = \langle {\it oi}, c\rangle$,
  where ${\it oi}$ is the object's identifier and $c$ is the class of the object,
  where $c\in C$.
  
\item ${\it OAT}$
  is a set of \emph{attribute values} ${\it atv}$,
  ${\it atv}$ = $\langle \langle \mathit{ati}$, $c$, $t\rangle$, $\langle\mathit{oi}$, $c\rangle$, $\mathit{vl}\rangle$,
where $\langle \mathit{ati}, c, t\rangle \in {\it AT}$,
$\langle oi, c\rangle\in {\it OC}$, and ${\it vl}$ is a value of the type $t$.
The attribute value ${\it atv}$
denotes the value ${\it vl}$ of the attribute  $\langle \mathit{ati}, c, t\rangle$
of the object  $\langle\mathit{oi}$, $c\rangle$.

\item $\mathit{OAS}$ is a set of  \emph{association links} ${\it asl}$,
${\it asl}= \langle\langle {\it asi}, \mathit{ase}_{\rm l}$, $c_{\rm l}$, $\mathit{ase}_{\rm r}$, $c_{\rm r}\rangle$,
 $\langle \mathit{oi}_{\rm l}$, $c_{\rm l}\rangle$, 
 $\langle \mathit{oi}_{\rm r}$, $c_{\rm r}\rangle\rangle$,
where 
$\langle {\it asi}, \mathit{ase}_{\rm l}$, $c_{\rm l}$,
$\mathit{ase}_{\rm r}$, $c_{\rm r}\rangle\in {\it AS}$,
$\langle oi_{\rm l}, c_{\rm l}\rangle\in {\it OC}$, and
$\langle oi_{\rm r}, c_{\rm r}\rangle\in {\it OC}$.
The association link ${\it asl}$
denotes that there is a link of the
association $\langle {\it asi}, \mathit{ase}_{\rm l}$, $c_{\rm l}$,
$\mathit{ase}_{\rm r}$, $c_{\rm r}\rangle$
between the objects
$\langle oi_{\rm l}, c_{\rm l}\rangle$ and
$\langle oi_{\rm r}, c_{\rm r}\rangle$, where the latter
stands at the end $\mathit{ase}_{\rm r}$
and the former stands at the end $\mathit{ase}_{\rm l}$.
\end{itemize}
\end{definition}

Let $o$ be an object. We denote by 
${\rm id}(o)$ the  identifier of the object $o$.
We assume that
every object is identified by a unique integer number.
 %
Let $\mathcal{D}$ be a data model. We denote by ${\rm Inst}(\mathcal{D})$ 
the set of 
instances of $\mathcal{D}$.

\section{The mapping OCL2MSFOL}
\label{ocl2msfol:app}
\subsubsection*{Mapping data models to MSFOL theories}

\cite{DaniaC16} defines a mapping  
${\rm o2f}_{\rm data}()$ 
from data models  to  MSFOL theories.
Let $\mathcal{D}$ be a data model. In a nutshell, ${\rm o2f}_{\rm data}(\mathcal{D})$ contains:
\begin{itemize}
\item The sorts {\tt Int} and {\tt String}, whose intended meaning is
to represent the integer numbers and the strings.
\item The constants {\tt nullInt}, {\tt nullString}, {\tt invalInt,} and {\tt inval\-String}, 
whose intended meaning is to represent {\tt null} and
{\tt invalid} for integers and strings.
\item The sort {\tt Classifier}, whose intended meaning is to represent all the objects in an instance of $\mathcal{D}$, as well as
{\tt null} and {\tt invalid} for objects.
\item For each class $c$ in $\mathcal{D}$, a unary predicate $c$, whose intended meaning is to define the objects of the class $c$
in an instance of $\mathcal{D}$.
\item For each attribute ${\it att}$ in $\mathcal{D}$, a function ${\it att}$, 
whose intended meaning is to define the values of the attribute
${\it att}$ in the objects in an instance of $\mathcal{D}$.
\item For each binary association ${\it as}$ in $\mathcal{D}$ 
with association-ends ${\it ase}_{\rm l}$ and ${\it ase}_{\rm r}$, 
a binary predicate  ${\it as}$, whose intended meaning
is to define the links via the association ${\it as}$  between objects in
 an instance of $\mathcal{D}$.
\item The axioms that constrain the meaning of the aforementioned sorts, constants, predicates, and functions.
\end{itemize}

Based on the definition of ${\rm o2f}_{\rm data}()$, we  define
the following mapping ${\rm intr}()$ from object models to MSFOL 
interpretations.
Let $\mathcal{D}$ be a data model and let $\mathcal{O}$ be
an instance of $\mathcal{D}$. Then,  ${\rm intr}(\mathcal{O})$ is the following
interpretation of  ${\rm o2f}_{\rm data}(\mathcal{D})$:
\begin{itemize}
\item The sort {\tt Int}  contains the integer numbers and the strings. The constants {\tt nullInt} and {\tt invalInt} are assigned two arbitrary (but different) integer numbers.
\item The sort {\tt String}  contains the strings. The constants {\tt null\-String} and {\tt invalString} are assigned two arbitrary (but different) strings.
\item The sort {\tt Classifier} contains the set of objects in $\mathcal{O}$
plus two distinguished elements, which are the
interpretations of the constants {\tt nullClassifier} and {\tt invalClassifier}.
\item For each class $c$ in $\mathcal{D}$, the predicate $c$ is assigned 
the set of objects  in $\mathcal{O}$ of  class $c$.
\item For each attribute ${\rm att}$ in $\mathcal{D}$, the function ${\it att}$ assigns
to each object in $\mathcal{O}$ the value of its attribute  ${\it att}$.
\item For each binary association ${\it as}$ in $\mathcal{D}$ with association-ends
${\rm as}_{\rm l}$ and ${\rm as}_{\rm r}$, the binary predicate ${\it as}$ is 
assigned all the pairs of objects linked through the association ${\it as}$ in 
$\mathcal{O}$.
\end{itemize}


\subsubsection*{Mapping OCL to MSFOL formulas}
Based on the definition of the mapping
${\rm o2f}_{\rm data}()$,
\cite{DaniaC16} defines a mapping
 from OCL Boolean
expressions  to MSFOL formulas. 
More precisely, it defines
 four mappings, namely, ${\rm o2f}_{\rm true}()$,
${\rm o2f}_{\rm false}()$, ${\rm o2f}_{\rm null}()$, and ${\rm o2f}_{\rm inval}()$, formalising, respectively, when an OCL Boolean expression evaluates to ${\tt true}$, ${\tt false}$,  ${\tt null}$,
or ${\tt invalid}$. 
These mappings are defined recursively over the
structure of OCL expressions. The following example
shows the recursive definition of these mappings.

\begin{example}
Consider the Boolean expression:
\begin{tabbing}
${\tt Student.allInstances()}\rightarrow{}{\tt notEmpty()}$.
\end{tabbing}
Then, according to the definition of the mapping
${\rm o2f}_{\rm true}()$ in~\cite{DaniaC16}:
\begin{tabbing}
${\rm o2f}_{\rm true}({\tt Student.allInstances()}\rightarrow{}{\tt notEmpty()})$\\
\quad = $\exists(x).({\rm o2f}_{\rm eval}({\tt Student.allInstances()})(x)$\\
\qquad $\wedge \neg({\rm o2f}_{\rm inval}({\tt Student.allInstances()})))$\\
\quad $= \exists(x).({\rm o2f}_{\rm eval}({\tt Student.allInstances()})(x))$
\end{tabbing}
where $x$ is a variable of type {\tt Classifier}.
\end{example}

Notice that, in the recursive case, when the subexpression is
a non-Boolean type, an auxiliary mapping 
${\rm o2f}_{\rm eval}()$ is called.
The mapping ${\rm o2f}_{\rm eval}()$ builds upon the 
mapping ${\rm o2f}_{\rm data}()$ from data models to MSFOL theories.
\cite{DaniaC16} distinguishes three  classes of non-Boolean expressions. 
The first class is formed by variables and by expressions
that denote primitive values and objects. Expressions denoting primitive values and objects are basically the literals (integers or strings), the arithmetic expressions, the expressions
denoting operations on strings, and the dot-expressions for
attributes.
Variables are mapped to MSFOL variables of the appropriate
sort. Expressions denoting primitive values and objects are
mapped by ${\rm o2f}_{\rm eval}()$ following the definition of the mapping
${\rm o2f}_{\rm data}()$. The output of the mapping ${\rm o2f}_{\rm eval}()$ for this first class
of non-Boolean expressions is always an MSFOL term.

\begin{example} Consider the non-Boolean expression:
{\tt p.age},
where {\tt p} is a variable of type {\tt Student}. Then,
 according to the definition of the mapping
${\rm o2f}_{\rm eval}()$ in~\cite{DaniaC16}:
\begin{tabbing}
${\rm o2f}_{\rm eval}({\tt p.age}) = {\tt age}({\rm o2f}_{\rm eval}({\tt p})) = {\tt age}({\tt p})$
\end{tabbing}
where {\tt p} is a variable of sort {\tt Classifier}.
\end{example}

The second class of non-Boolean expressions is formed
by the expressions that define sets. These expressions are
basically the {\tt allInstances}-expressions, the {\tt select} and 
{\tt reject}-expressions, the {\tt including} and 
{\tt excluding}-expressions, the {\tt intersection} and {\tt union}-expressions, 
and the {\tt collect}-expressions.
Each expression ${\it expr}$ in this class is mapped by ${\rm o2f}_{\rm eval}()$ to
a new predicate, denoted as $\ulcorner {\it expr}\urcorner$. 
This predicate formalises the set defined by the expression ${\it expr}$ and its definition
is generated by calling a mapping ${\rm o2f}_{\rm def\_ c}()$, which is also
defined in~\cite{DaniaC16}.

\begin{example}
Consider the non-Boolean expression:
\begin{tabbing}
${\tt Student.allInstances()}\rightarrow{}{\tt select(s|s.age.oclIsUndefined())}.$
\end{tabbing}
Then, according to the definition of the mapping
${\rm o2f}_{\rm eval}()$ in~\cite{DaniaC16}:
\begin{tabbing}
${\rm o2f}_{\rm eval}({\tt Student.allInstances()}$\\
\qquad $\rightarrow{}{\tt select(p|p.age.oclIsUndefined())})$\\
$= \ulcorner{\tt Student.allInstances()}$\\
\qquad $\rightarrow{}{\tt select(p|p.age.oclIsUndefined())}\urcorner$
\end{tabbing}
where the new predicate
\begin{tabbing}
$\ulcorner{\tt Student.allInstances()}$\\
\quad $\rightarrow{\tt select(p\mid{}p.age.oclIsUndefined())}\urcorner$
\end{tabbing}
is defined by ${\rm o2f}_{\rm def\_c}()$ as follows:
\begin{tabbing}
$\forall(s).(
\ulcorner{\tt Student.allInstances()}$\\
\qquad\quad $\rightarrow{\tt select(p|p.age.oclIsUndefined())}\urcorner
(s)$\\
\quad $\Leftrightarrow ({\rm o2f}_{\rm eval}({\tt Student.allInstances()})(s)$\\
\qquad\quad $\wedge\ {\rm o2f}_ {\rm true}({\tt s.age.oclIsUndefined()})))$\\
$=$\\
$\forall(s).(\ulcorner{\tt Student.allInstances()}\urcorner$\\
\qquad\quad $\rightarrow{}{\tt select(p|p.age.oclIsUndefined())}\urcorner(s)$\\
\quad $\Leftrightarrow (\ulcorner{\tt Student.allInstances()}\urcorner(s)$\\
\quad\qquad$\wedge\ {\rm o2f}_{\rm eval}({\tt s.age}) = {\tt nullInt}$\\
\quad\qquad $\vee (s = {\tt nullClassifier} \vee s = {\tt invalClassifier})))$\\
$=$\\
$\forall(s).(\ulcorner{\tt Student.allInstances()}\urcorner$\\
\qquad \quad $\rightarrow{}{\tt select(p|p.age.oclIsUndefined())}\urcorner(s)$\\
\quad $\Leftrightarrow (\ulcorner{\tt Student.allInstances()}\urcorner(s)$\\
\qquad $\wedge\ {\tt age(s)} = {\tt nullInt}$\\
\qquad $\vee ({\tt s} = {\tt nullClassifier} \vee  {\tt s} = {\tt invalClassifier})))$
\end{tabbing}
where the new predicate $\ulcorner{\tt Student.allInstances()}\urcorner$ 
is defined by ${\rm o2f}_{{\rm def}\_{c}}()$ as follows:
\begin{tabbing}
$\forall(s).(\ulcorner{\tt Student.allInstances()}\urcorner \Leftrightarrow {\tt Student}(s))$
\end{tabbing}
where $s$ is a variable of type {\tt Classifier}.
\end{example}

The third class of non-Boolean expressions is formed by
the expressions that distinguish an element from a set. These
expressions are, basically, the {\tt any}, {\tt max}, and {\tt min}-expressions.
Each expression ${\it expr}$ in this class is mapped by ${\rm o2f}_{\rm eval}$() to
a new function, denoted as $\ulcorner {\it expr}\urcorner$, which represents 
the element referred to by ${\it expr}$. The axioms defining
$\ulcorner {\it expr}\urcorner$ are generated by calling a mapping
 ${\rm o2f}_{\rm def\_ o}()$, which is also
defined in~\cite{DaniaC16}.

\cite{DaniaC16} denotes by ${\rm o2f}_{\rm def}({\it expr})$ the set of axioms that result from applying ${\rm o2f}_{\rm def\_c}()$ and ${\rm o2f}_{\rm def\_o}()$ to the
corresponding non-Boolean subexpression in ${\it expr}$. 
In particular, for each literal integer $i$ and literal string
${\it st}$ in ${\it expr}$, ${\rm o2f}_{\rm def\_o}()$ generates the following axioms:
\begin{tabbing}
${\rm o2f}_{\rm def\_o}(i) = \neg(i = {\tt nullInt}) \wedge \neg(i = {\tt invalInt}).$\\
${\rm o2f}_{\rm def\_o}({\it st}) = \neg({\it st} = {\tt nullString}) \wedge \neg({\it st} = {\tt invalString})$
\end{tabbing}

\section{The mapping OCL2PSQL}
\label{ocl2psql:app}
\subsubsection*{Mapping data models to SQL schemata}

\cite{BaoC19} introduces a mapping, called OCL2PSQL, 
from OCL to SQL. 
It contains two parts: first, a mapping 
from data models to SQL  schemata, 
and then a mapping  from OCL expressions to pure SQL select statements.
The mapping from data models 
to SQL schemata, which we denoted as ${\rm o2s}()$, is 
the usual OR mapping (classes
are mapped to tables, attributes to columns, and many-to-many associations to
tables with appropriate foreign-keys).
Let $\mathcal{D} = \langle C, {\it AT}, {\it AS}\rangle$ be a 
data model. In a nutshell, ${\rm o2s}(\mathcal{D})$
contains the following statements:
\begin{itemize}
\item For each class $c\in C$, 
\begin{tabbing}
{\tt >} {\tt CREATE TABLE}\ $\ulcorner c\urcorner$\ {\tt (} $c${\tt\_id int PRIMARY KEY)}
\end{tabbing}
\item For each attribute ${\it att} = \langle{\it ati}, c, t\rangle$ in ${\it AT}$,
\begin{tabbing}
{\tt >} {\tt ALTER TABLE}\ $\ulcorner c\urcorner$\ {\tt ADD COLUMN}\ ${\it att}$\
${\rm SqlType}(t)$
\end{tabbing}
where:
\begin{itemize}
\item if $t$ = {\tt Integer}, then ${\rm SqlType}(t) = 
{\tt int}$;
\item if $t$ = {\tt String}, then ${\rm SqlType}(t) 
= {\tt varchar}$;
\item if $t \in  C$, then ${\rm SqlType}(t) = {\tt int}$.
\end{itemize}
Moreover, if $t \in C$, then
\begin{tabbing}
{\tt >} {\tt ALTER TABLE}\ $\ulcorner c\urcorner$\ {\tt ADD FOREIGN KEY fk\_}$c${\tt\_}{\it ati}(${\it att}$)\\
{\tt >} \quad {\tt REFERENCES}\ $\ulcorner t\urcorner$($t${\tt\_id});
\end{tabbing}
\item For each association ${\it as} = \langle {\it asi}, \mathit{ase}_{\rm l}$, $c_{\rm l}$,
  $\mathit{ase}_{\rm r}$, $c_{\rm r}\rangle$ in 
  ${\it AS}$,
 \begin{tabbing} 
{\tt >} {\tt CREATE TABLE $\ulcorner {\it as}\urcorner$ (}
${\it ase}_{\rm l}$\ {\tt int,}
${\it ase}_{\rm r}$\ {\tt int,}\\
{\tt >} \quad {\tt FOREIGN KEY fk\_}$c_{\rm l}{\tt\_}{\it ase}_{\rm l}(
{\it ase}_{\rm l})$ {\tt REFERENCES}\ $\ulcorner c_{\rm l}\urcorner( c_{\rm l}{\tt\_id})${\tt ,}\\
{\tt >} \quad {\tt FOREIGN KEY fk\_}$c_{\rm r}{\tt\_}
{\it ase}_{\rm r}(
{\it ase}_{\rm r})$ {\tt REFERENCES}\ $\ulcorner c_{\rm r}\urcorner( c_{\rm r}{\tt\_id}))$.
\end{tabbing}
\end{itemize}

\cite{BaoC19} also defines a mapping ${\rm o2s}_{\rm inst}()$ from instances of data models
to instances of SQL databases.
Let $\mathcal{D}  = \langle C , {\it AT}, {\it AS}\rangle$ be a data model. Let 
$\mathcal{O} = \langle {\it OC}, 
{\it OAT},
{\it OAS}\rangle$ be an instance of $\mathcal{D}$. Then ${\rm o2s}_{\rm inst}(\mathcal{O})$ 
is defined as follows:
\begin{itemize}
\item For each  $o = \langle {\it oi}, c\rangle \in  
{\it OC}$,
\begin{tabbing}
{\tt >} {\tt INSERT INTO}\ $\ulcorner c\urcorner$\ {\tt (}$c${\tt\_id})\ {\tt VALUES}\ {\tt(}${\rm id}(o)${\tt )}
\end{tabbing}

\item For each $\langle {\it att},
(o, c), v\rangle\in {\it OAT}$,
\begin{tabbing}
{\tt >} {\tt UPDATE}\ $\ulcorner c\urcorner$\ 
{\tt SET}\ ${\it att}$\ {\tt =}\ $v$\ {\tt WHERE} 
$c${\tt{\_id}}\ {\tt =}\ ${\rm id}(o)$
\end{tabbing}
\item For each $\langle\langle {\it as}, {\it ase}_{\rm l}, 
c_{\rm l},  {\it ase}_{\rm r}, c_{\rm r}\rangle,
{\it o}_{\rm l},
{\it o}_{\rm r}\rangle
\in {\it OAS}$,
\begin{tabbing}
{\tt >} {\tt INSERT INTO}\ $\ulcorner{\it as}\urcorner$ \ 
{\tt (}${\it ase}_{\rm l}${\tt ,} ${\it ase}_{\rm r}${\tt )} {\tt VALUES}\ {\tt (}${\rm id}(o_{\rm l})${\tt ,} 
${\rm id}(o_{\rm r})${\tt )}
\end{tabbing}
\end{itemize}

\section{The mapping SQL2MSFOL}
\label{sql2msfol:app}

\subsubsection*{Mapping SQL schemata to MSFOL theories}

Let $\mathcal{D}$ be a data model.
We recall that our mapping SQL2MSFOL
assumes that the SQL implementations of the 
OCL constraints in the context of the data model $\mathcal{D}$
 are  select-statements in the context of the SQL schema 
${\rm o2s}(\mathcal{D})$
generated 
by the mapping OCL2PSQL~\cite{BaoC19} 
in Appendix~\ref{ocl2psql:app}.

\emph{Notation.} In what follows, for any class ${\it c}$
in the contextual model,
we use ${\rm index}_{\ulcorner c\urcorner}(\_)$ 
to denote the name
of a new unary predicate. 
Similarly, for any association ${\it as}$ in the contextual model,
we use ${\rm index}_{\ulcorner{\it as}\urcorner}(\_)$ to denote the name
of a new unary predicate. In what follows, 
the variable $c, c'$ are of sort ${\tt Classifier}$,
and the variable $x, y, z$ are of sort ${\tt Int}$.
Also, ${\rm id}()$ is  an uninterpreted function 
from the sort ${\tt Int}$ to the sort ${\tt Classifier}$,
and ${\rm left}$, and ${\rm right}$ are uninterpreted functions 
from the sort ${\tt Int}$ to the sort ${\tt Int}$.

For each class $C$ in the contextual model, 
the mapping ${\rm s2f}()$ generates the following axioms:

\begin{tabbing}
$\forall(x)({\rm index}_{\ulcorner c\urcorner}(x) \Rightarrow \exists(c)(C(c) \land c = {\rm id}(x)))$.
\end{tabbing}
\begin{tabbing}
$\forall(c)(C(c) \Rightarrow \exists(x)({\rm index}_{\ulcorner{\it c}\urcorner}(x) \land c = {\rm id}(x)))$.
\end{tabbing}
\begin{tabbing}
$\forall(x, y)(({\rm index}_{\ulcorner{\it c}\urcorner}(x) \wedge
{\rm index}_{\ulcorner{\it c}\urcorner}(y) \wedge
x \not= y) \Rightarrow ({\rm id}(x) \not= {\rm id}(y)))$.
\end{tabbing}
\begin{tabbing}
$\forall(x)({\rm index}_{\ulcorner{\it C}\urcorner}(x) \Rightarrow
{\tt val}_{\ulcorner{\it c}\urcorner}({\ulcorner{\it c}\urcorner}{\tt \_id}, x) 
= {\rm id}(x))$.
\end{tabbing}

\noindent and for each attribute ${\it att}$ in $C$, 

\begin{tabbing}
$\forall(x)({\rm index}_{\ulcorner{\it c}\urcorner}(x) \Rightarrow
{\tt val}_{\ulcorner{\it c}\urcorner}({\it att}, x) = {\it att}({\rm id}(x)))$.
\end{tabbing}

Moreover, for each association ${\it as}$,
with association-ends ${\it ase}_{\rm l}$ and ${\it ase}_{\rm r}$ 
in the contextual model,,
${\rm s2f}()$ generates the following axioms: 

\begin{tabbing}
$\forall(x, y)({\rm index}_{\ulcorner{\it as}\urcorner}(x) 
\wedge {\rm index}_{\ulcorner{\it as}\urcorner}(y) \wedge x \not= y$\\
\quad $\Rightarrow
\neg({\rm left}(x) = {\rm left}(y) \wedge
{\rm right}(x) = {\rm right}(y)))$.
\end{tabbing}
\begin{tabbing}
$\forall(c, c')({\it as}(c, c')$\\
\quad $ \Rightarrow \exists(x)({\rm index}_{\rm \ulcorner {\it as}\urcorner}(x) \wedge {\rm id}({\rm left}(x)) = c \wedge {\rm id}({\rm right}(x)) = c')$.
\end{tabbing}
\begin{tabbing}
$\forall(x)({\rm index}_{\rm \ulcorner {\it as}\urcorner}(x)$\\
\quad $ \Rightarrow \exists(c, c')({\it as}(c, c') \wedge {\rm id}({\rm left}(x)) = c
\wedge {\rm id}({\rm right}(x)) = c'$.
\end{tabbing}
\begin{tabbing}
$\forall(x)({\rm index}_{\ulcorner{\it as}\urcorner}(x) \Rightarrow
{\tt val}_{\ulcorner{\it as}\urcorner}({\it ase}_{\rm l}, x) 
= {\rm id}({\rm left}(x)))$.
\end{tabbing}
\begin{tabbing}
$\forall(x)({\rm index}_{\ulcorner{\it as}\urcorner}(x) \Rightarrow
{\tt val}_{\ulcorner{\it as}\urcorner}({\it ase}_{\rm r}, x) 
= {\rm id}({\rm right}(x)))$.
\end{tabbing}

\subsubsection*{Mapping select-statements}
Currently, our mapping ${\rm s2f}()$ covers the  SQL patterns
below, where ${\it fromitem}$ is either a \emph{table} 
of a \emph{subselect}. 
We do not consider correlated subqueries.
\begin{itemize}
\item {\tt SELECT {\it selectitems}}.
\item {\tt SELECT {\it selectitems} FROM {\it fromitem}}.
\item {\tt SELECT {\it selectitems} FROM {\it fromitem} WHERE {\it bexpr}}.
\item {\tt SELECT {\it selectitems} FROM {\it fromitem} JOIN {\it fromitem'}}.
\item {\tt SELECT {\it selectitems} FROM {\it fromitem}\\ JOIN {\it fromitem'} ON {\it bexpr}}.
\item {\tt SELECT {\it selectitems} FROM {\it fromitem}\\ JOIN {\it fromitem'} ON {\it bexpr} WHERE {\it bexpr'}}.
\end{itemize}

\emph{Notation.} In what follows, for any select-statement ${\it sel}$
we use ${\rm index}_{\it sel}(\_)$ to denote the name
of a new unary predicate. 
Also,  for any from-expression ${\it fromitem}$
we use ${\rm index}_{\it fromitem}(\_)$ to denote the name
of a new unary predicate. 
And, similarly, for any select-statement ${\it sel}$,
and any expression ${\it expr}$, 
we use ${\rm val}_{\it sel}({\it expr}, \_)$ to denote the name
of a new unary function. 

For each (sub)select statement, ${\rm s2f}()$ generates the  following axioms: 
\paragraph{{\bf Case:} {\it sel} $\coloneqq$ {\tt SELECT {\it selectitems}}}
\begin{tabbing}
$\exists(x)({\rm index}_{\it sel}(x) 
\wedge \forall(y)(y \not= x \Rightarrow 
\neg({\rm index}_{\it sel}(y)))).$
\end{tabbing}
\paragraph{{\bf Case:} {\it sel} $\coloneqq$ {\tt SELECT {\it selectitems} FROM {\it fromitem}}.} 
\begin{tabbing}
$\forall(x)({\rm index}_{\it sel}(x) \iff {\rm index}_{\it fromitem}(x))$.
\end{tabbing}

Moreover, for each column
${\it fromitem}.{\it col}$ in ${\it selectitems}$,
\begin{tabbing}
$\forall(x)(
{\rm index}_{\it sel}(x)$ $\Rightarrow {\tt val}_{\it sel}({\it fromitem}.{\it col}, x)
= {\tt val}_{\it fromitem}({\it col}, x))$.
\end{tabbing}
\paragraph{{\bf Case:} {\it sel} $\coloneqq$ {\tt SELECT {\it selectitems} FROM {\it fromitem} WHERE {\it expr}}.}
\begin{tabbing}
$\forall(x)({\rm index}_{\it sel}(x)$\\
\quad $\iff {\rm index}_{\it fromitem}(x) \land\ {\tt val}_{\it fromitem}({\it expr}, x) = \underline{{\tt TRUE}})$.
\end{tabbing}

Moreover, for each column
${\it fromitem}.{\it col}$ in ${\it selectitems}$,
\begin{tabbing}
$\forall(x)(
{\rm index}_{\it sel}(x)$ $\Rightarrow {\tt val}_{\it sel}({\it fromitem}.{\it col}, x)
= {\tt val}_{\it fromitem}({\it col}, x))$.
\end{tabbing}

\paragraph{{\bf Case:} {\it sel} $\coloneqq$ {\tt SELECT {\it selectitems} FROM {\it fromitem} JOIN {\it fromitem'}}.}
The predicate
${\rm index}_{\it join}(\_)$  specifies
the indices of the intermediate table resulting from
joining ${\it fromitem}$ with ${\it fromitem'}$.
\begin{tabbing}
$\forall(x)({\rm index}_{\it sel}(x) \iff {\rm index}_{\it join}(x))$.
\end{tabbing}
\begin{tabbing}
$\forall(x, y)({\rm index}_{\it join}(x) 
\wedge {\rm index}_{\it join}(y) \wedge x \not= y$\\
\quad $\Rightarrow
\neg({\rm left}(x) = {\rm left}(y) \wedge
{\rm right}(x) = {\rm right}(y)))$.
\end{tabbing}
\begin{tabbing}
$\forall(x)({\rm index}_{\it join}(x) \Rightarrow \exists(y,z)({\rm index}_{\it fromitem}(y) $\\
\quad $\wedge\  {\rm index}_{\it fromitem'}(z) \wedge y = {\rm left}(x) \wedge z = {\rm right}(x)))$.
\end{tabbing}
\begin{tabbing}
$\forall(y,z)({\rm index}_{\it fromitem}(y) \land {\rm index}_{\it fromitem'}(z)$\\
\quad $\Rightarrow \exists(x)({\rm index}_{\it join}(x) \wedge y = {\rm left}(x) \wedge z = {\rm right}(x)))$.
\end{tabbing}

Moreover, for each column
${\it fromitem}.{\it col}$ in ${\it selectitems}$,
\begin{tabbing}
$\forall(x)(
{\rm index}_{\it sel}(x)$\\ 
\quad $\Rightarrow {\tt val}_{\it sel}({\it fromitem}.{\it col}, x)
= {\tt val}_{\it fromitem}({\it col}, {\rm left}(x)))$.
\end{tabbing}

Finally for each column
${\it fromitem'}.{\it col}$ in ${\it selectitems}$,
\begin{tabbing}
$\forall(x)(
{\rm index}_{\it sel}(x)$\\ 
\quad $\Rightarrow {\tt val}_{\it sel}({\it fromitem'}.{\it col}, x)
= {\tt val}_{\it fromitem'}({\it col}, {\rm right}(x)))$.
\end{tabbing}


\paragraph{{\bf Case:} {\it sel} $\coloneqq$ {\tt SELECT {\it selectitems} FROM {\it fromitem} JOIN {\it fromitem'} ON {\it bexpr}}.}
The predicate
${\rm index}_{\it join}(\_)$  specifies
the indices of the intermediate table resulting from
joining ${\it fromitem}$ with ${\it fromitem'}$ taking 
into account the on-clause ${\it bexpr}$.

\begin{tabbing}
$\forall(x)({\rm index}_{\it sel}(x) \iff {\rm index}_{\it join}(x)$\\ 
\quad $\land\ {\tt val}_{\it join}({\it bexpr}, x) = 
\underline{{\tt TRUE}})$.
\end{tabbing}
\begin{tabbing}
$\forall(x, y)({\rm index}_{\it join}(x) 
\wedge {\rm index}_{\it join}(y) \wedge x \not= y$\\
\quad $\Rightarrow
\neg({\rm left}(x) = {\rm left}(y) \wedge
{\rm right}(x) = {\rm right}(y)))$.
\end{tabbing}
\begin{tabbing}
$\forall(x)({\rm index}_{\it join}(x) \Rightarrow \exists(y,z)({\rm index}_{\it fromitem}(y) $\\
\quad $\wedge\ {\rm index}_{\it fromitem'}(z) \wedge y = {\rm left}(x) \wedge z = {\rm right}(x)))$.
\end{tabbing}
\begin{tabbing}
$\forall(y,z)({\rm index}_{\it fromitem}(y) \land {\rm index}_{\it fromitem'}(z)$\\
\quad $\Rightarrow \exists(x)({\rm index}_{\it join}(x) \wedge y = {\rm left}(x) \wedge z = {\rm right}(x)))$.
\end{tabbing}

Moreover, for each column
${\it fromitem}.{\it col}$ in ${\it selectitems}$,
\begin{tabbing}
$\forall(x)(
{\rm index}_{\it sel}(x) \Rightarrow {\tt val}_{\it sel}({\it fromitem}.{\it col}, x)
= {\tt val}_{\it fromitem}({\it col}, {\rm left}(x)))$.
\end{tabbing}

Finally for each column
${\it fromitem'}.{\it col}$ in ${\it selectitems}$,
\begin{tabbing}
$\forall(x)(
{\rm index}_{\it sel}(x)$\\ 
\quad $\Rightarrow {\tt val}_{\it sel}({\it fromitem'}.{\it col}, x)
= {\tt val}_{\it fromitem'}({\it col}, {\rm right}(x)))$.
\end{tabbing}

\paragraph{{\bf Case:} {\it sel} $\coloneqq$ {\tt SELECT {\it selectitems} FROM {\it fromitem} JOIN {\it fromitem'} ON {\it bexpr} WHERE {\it bexpr'}}.}

The predicate
${\rm index}_{\it join}(\_)$  specifies
the indices of the intermediate table resulting from
joining ${\it fromitem}$ with ${\it fromitem'}$ taking 
into account the on-clause ${\it bexpr}$ and the
where-clause ${\it bexpr'}$

%
\begin{tabbing}
$\forall(x)({\rm index}_{\it sel}(x) \iff {\rm index}_{\it join}(x)$\\
\quad $\land\ {\tt val}_{\it join}({\it bexpr}, x) = \underline{{\tt TRUE}} \land\ {\tt val}_{\it join}({\it bexpr'}, x) = \underline{{\tt TRUE}})$.
\end{tabbing}
\begin{tabbing}
$\forall(x, y)({\rm index}_{\it join}(x) 
\wedge {\rm index}_{\it join}(y) \wedge x \not= y$\\
\quad $\Rightarrow
\neg({\rm left}(x) = {\rm left}(y) \wedge
{\rm right}(x) = {\rm right}(y)))$.
\end{tabbing}
\begin{tabbing}
$\forall(x)({\rm index}_{\it join}(x) \Rightarrow \exists(y,z)({\rm index}_{\it fromitem}(y) $\\
\quad $\wedge\  {\rm index}_{\it fromitem'}(z) \wedge y = {\rm left}(x) \wedge z = {\rm right}(x)))$.
\end{tabbing}
\begin{tabbing}
$\forall(y,z)({\rm index}_{\it fromitem}(y) \land {\rm index}_{\it fromitem'}(z)$\\
\quad $\Rightarrow \exists(x)({\rm index}_{\it join}(x) \wedge y = {\rm left}(x) \wedge z = {\rm right}(x)))$.
\end{tabbing}

Moreover, for each column
${\it fromitem}.{\it col}$ in ${\it selectitems}$,
\begin{tabbing}
$\forall(x)(
{\rm index}_{\it sel}(x)$\\ 
\quad $\Rightarrow {\tt val}_{\it sel}({\it fromitem}.{\it col}, x)
= {\tt val}_{\it fromitem}({\it col}, {\rm left}(x)))$.
\end{tabbing}

Finally for each column
${\it fromitem'}.{\it col}$ in ${\it selectitems}$,
\begin{tabbing}
$\forall(x)(
{\rm index}_{\it sel}(x)$\\ 
\quad $\Rightarrow {\tt val}_{\it sel}({\it fromitem'}.{\it col}, x)
= {\tt val}_{\it fromitem'}({\it col}, {\rm right}(x)))$.
\end{tabbing}

\subsubsection*{Mapping expressions}

Currently, our mapping ${\rm s2f}()$ covers the following SQL expressions:\\

\begin{tabular}{llll}
${\it expr}$ & $\coloneqq$ & ${\tt TRUE} \mid {\tt FALSE} \mid {\tt NULL}$ & ({\rm Boolean literals}) \\ 
& $\mid$ & $ {\tt NULL} \dots \mid 0 \mid 1 \mid \dots$ & 
({\rm integer literals}) \\
& $\mid$ & ${\it var}$ & ({\rm variables}) \\
& $\mid$ & $c\_{\tt id} \mid {\it attribute}$ & ({\rm class ids and attributes}) \\
& $\mid$ & ${\it association\mbox{-}end}$ & ({\rm association-end}) \\
& $\mid$ & $\ominus {\it expr}$ & ({\rm unary logical ops}) \\
& $\mid$ & ${\it expr}_1 \oplus {\it expr}_2$ & 
({\rm binary logical ops}) \\
& $\mid$ & ${\it expr}_1 \otimes {\it expr}_2$ & 
({\rm binary comparison ops}) \\
& $\mid$ & \multicolumn{2}{l}{$\texttt{CASE WHEN {\it expr}}\ \texttt{THEN}\ {\it expr}_1\ {\tt ELSE}\ {\it expr}_2\ {\tt END}$} \\ 
& & & ({\tt CASE}{\rm -expression}) \\
& $\mid$ & ${\it expr}\ \texttt{IS NULL}$ & 
({\tt IS NULL}{\rm -expression}) \\
& $\mid$ & ${\tt EXISTS}\ {\it subselect}$ & 
({\tt EXISTS}{\rm -expression}) \\
& $\mid$ & ${\it subselect}$ & ({\rm single-valued subselect}) \\
\\
\end{tabular}

Let ${\it sel}$ be a (sub)select statement.
Then, for  each (sub) expression ${\it expr}$ in ${\it sel}$, ${\rm s2f}()$ generates
the following axioms.

\paragraph{{\bf Case:} ${\it expr} \coloneqq {\tt TRUE} \mid {\tt FALSE} \mid {\tt NULL}$.}
\begin{tabbing}
$\forall(x)({\rm index}_{\it sel}(x) \Rightarrow {\rm val}_{\it sel}({\tt TRUE}, x) 
= \underline{{\tt TRUE}})$.\\
$\forall(x)({\rm index}_{\it sel}(x) \Rightarrow {\rm val}_{\it sel}({\tt FALSE}, x) 
= \underline{{\tt FALSE}})$.\\
$\forall(x)({\rm index}_{\it sel}(x) \Rightarrow {\rm val}_{\it sel}({\tt NULL}, x) = 
\underline{{\tt NULL}}).$
\end{tabbing}

\paragraph{{\bf Case:} ${\it expr} \coloneqq \dots \mid -1 \mid 0 \mid {\tt NULL} \mid 1 \mid \dots$.}
\begin{tabbing}
$\forall(x)({\rm index}_{\it sel}(x) \Rightarrow {\rm val}_{\it sel}({\tt NULL}, x) 
= {\tt nullInt})$.
\end{tabbing}
For every integer number $i$, 
\begin{tabbing}
$\forall(x)({\rm index}_{\it sel}(x) \Rightarrow {\rm val}_{\it sel}(i, x) = i$.
\end{tabbing}


\paragraph{{\bf Case:} ${\it expr}\coloneqq 
{\it var}$.} 

\begin{tabbing}
$\forall(x).({\rm index}_{\it sel}(x) \Rightarrow ({\rm val}_{\it sel}({\it var}, x)= {\it var}))$.
\end{tabbing}

\paragraph{{\bf Case:} ${\it expr}\coloneqq c\_{\tt id} \mid 
{\it attribute} \mid {\it association\mbox{-}end}$.} 
The corresponding axioms are introduced
in \emph{Mapping SQL schemata to MSFOL theories}
above.

\paragraph{{\bf Case:} ${\it expr} \coloneqq {\tt NOT}\ {\it expr'}$.}
\begin{tabbing}
$\forall(x).({\rm index}_{\it sel}(x) \Rightarrow ({\rm val}_{\it sel}({\it expr}, x)= \underline{{\tt TRUE}}$\\
\quad $\iff
{\rm val}_{\it sel}({\it expr'}, x)= \underline{{\tt FALSE}}))$.\\
$\forall(x).({\rm index}_{\it sel}(x) \Rightarrow ({\rm val}_{\it sel}({\it expr}, x)= \underline{{\tt FALSE}}$\\ 
\quad $\iff
{\rm val}_{\it sel}({\it expr'}, x)= \underline{{\tt TRUE}}))$.\\
$\forall(x).({\rm index}_{\it sel}(x) \Rightarrow ({\rm val}_{\it sel}({\it expr}, x)= \underline{{\tt NULL}}$\\ 
\quad $\iff
{\rm val}_{\it sel}({\it expr'}, x)= \underline{{\tt NULL}}))$.
\end{tabbing}

\paragraph{{\bf Case:} ${\it expr} \coloneqq {\it expr}_1\ {\tt AND}\ {\it expr}_2$.}
\begin{tabbing}
$\forall(x).({\rm index}_{\it sel}(x) \Rightarrow {\rm val}_{\it sel}({\it expr}, x)= \underline{{\tt TRUE}}$\\
\quad $\iff
{\rm val}_{\it sel}({\it expr}_1, x)= \underline{{\tt TRUE}}
\land {\rm val}_{\it sel}({\it expr}_2, x)= \underline{{\tt TRUE}})$.\\
$\forall(x).({\rm index}_{\it sel}(x) \Rightarrow {\rm val}_{\it sel}({\it expr}, x)= \underline{{\tt FALSE}}$\\
\quad $\iff
{\rm val}_{\it sel}({\it expr}_1, x)= \underline{{\tt FALSE}}
\lor {\rm val}_{\it sel}({\it expr}_2, x)= \underline{{\tt FALSE}})$.\\
$\forall(x).({\rm index}_{\it sel}(x) \Rightarrow ({\rm val}_{\it sel}({\it expr}, x)= \underline{{\tt NULL}}$\\
\quad $\iff
({\rm val}_{\it sel}({\it expr}_1, x)= \underline{{\tt NULL}} \land
 {\rm val}_{\it sel}({\it expr}_2, x)= \underline{{\tt NULL}})$\\
\quad $\lor ({\rm val}_{\it sel}({\it expr}_1, x)= 
\underline{{\tt NULL}} \land
 {\rm val}_{\it sel}({\it expr}_2, x)= \underline{{\tt TRUE}})$\\
\quad $\lor ({\rm val}_{\it sel}({\it expr}_1, x)= 
\underline{{\tt TRUE}} \land
 {\rm val}_{\it sel}({\it expr}_2, x)= \underline{{\tt NULL}})))$.
\end{tabbing}

\paragraph{{\bf Case:} ${\it expr} \coloneqq {\it expr}_1\ {\tt OR}\ {\it expr}_2$.}
\begin{tabbing}
$\forall(x).({\rm index}_{\it sel}(x) \Rightarrow {\rm val}_{\it sel}({\it expr}, x)= \underline{{\tt TRUE}}$\\
\quad $\iff
{\rm val}_{\it sel}({\it expr}_1, x)= \underline{{\tt TRUE}}
\lor {\rm val}_{\it sel}({\it expr}_2, x)= \underline{{\tt TRUE}})$.\\
$\forall(x).({\rm index}_{\it sel}(x) \Rightarrow {\rm val}_{\it sel}({\it expr}, x)= \underline{{\tt FALSE}}$\\
\quad $\iff
{\rm val}_{\it sel}({\it expr}_1, x)= \underline{{\tt FALSE}}
\land {\rm val}_{\it sel}({\it expr}_2, x)= \underline{{\tt FALSE}})$.\\
$\forall(x).({\rm index}_{\it sel}(x) \Rightarrow ({\rm val}_{\it sel}({\it expr}, x)= \underline{{\tt NULL}}$\\
\quad $\iff
({\rm val}_{\it sel}({\it expr}_1, x)= \underline{{\tt NULL}} \land
 {\rm val}_{\it sel}({\it expr}_2, x)= \underline{{\tt NULL}})$\\
\quad $\lor ({\rm val}_{\it sel}({\it expr}_1, x)= 
\underline{{\tt NULL}} \land
 {\rm val}_{\it sel}({\it expr}_2, x)= \underline{{\tt FALSE}})$\\
\quad $\lor ({\rm val}_{\it sel}({\it expr}_1, x)= 
\underline{{\tt FALSE}} \land
 {\rm val}_{\it sel}({\it expr}_2, x)= 
 \underline{{\tt NULL}})))$.
\end{tabbing}

\paragraph{{\bf Case:} ${\it expr} \coloneqq {\it expr'}\ \texttt{IS NULL}$.}
Let $t$ be the type of ${\it expr'}$. Then,
\begin{tabbing}
$\forall(x).({\rm index}_{\it sel}(x) \Rightarrow ({\rm val}_{\it sel}({\it expr}, x)= \underline{{\tt TRUE}}$\\
\quad $\iff {\rm val}_{\it sel}({\it expr'}, x)= {\rm nullOf}(t)))$.\\
$\forall(x).({\rm index}_{\it sel}(x) \Rightarrow ({\rm val}_{\it sel}({\it expr}, x)= \underline{{\tt FALSE}}$\\
\quad $\iff {\rm val}_{\it sel}({\it expr'}, x) \not= {\rm nullOf}(t)))$.
\end{tabbing}
where ${\rm nullOf}({\tt Bool}) = \underline{{\tt NULL}}$, 
${\rm nullOf}({\tt Integer}) = {\tt nullInt}$, and
${\rm nullOf}({\tt String}) = {\tt nullString}$; 

\paragraph{{\bf Case:} ${\it expr} \coloneqq {\it expr}_1 \otimes {\it expr}_2$.} 
\begin{tabbing}
$\forall(x).({\rm index}_{\it sel}(x) \Rightarrow ({\rm val}_{\it sel}({\it expr}, x) = \underline{{\tt TRUE}}$ \\
\quad $\iff \neg({\rm val}_{\it sel}({{\it expr}_1}, x) = {\rm nullOf}(t))$ \\
\quad $\land \neg({\rm val}_{\it sel}({{\it expr}_2},x) = {\rm nullOf}(t))\ \land\  {\rm val}_{\it sel}({{\it expr}_1}, x) \otimes {\rm val}({{\it expr}_2}, x)))$. \\
$\forall(x).({\rm index}_{\it sel}(x) \Rightarrow ({\rm val}_{\it sel}({\it expr}, x) = \underline{{\tt FALSE}}$ \\
\quad $\iff \neg({\rm val}_{\it sel}({{\it expr}_1}, x) = {\rm nullOf}(t))$ \\
\quad $\land \neg({\rm val}_{\it sel}({{\it expr}_2},x) = {\rm nullOf}(t))\ \land\  \neg({\rm val}_{\it sel}({{\it expr}_1}, x) \otimes {\rm val}({{\it expr}_2}, x))))$. \\
$\forall(x).({\rm index}_{\it sel}(x) \Rightarrow ({\rm val}_{\it sel}({\it expr}, x) = \underline{{\tt NULL}}$ \\
\quad $\iff {\rm val}_{\it sel}({{\it expr}_1}, x) = {\rm nullOf}(t)\ \vee {\rm val}_{\it sel}({{\it expr}_2}, x) = {\rm nullOf}(t)))$. 
\end{tabbing}

\paragraph{{\bf Case:} ${\it expr} \coloneqq \texttt{CASE WHEN {\it expr'} THEN}\ {\it expr}_1\ {\tt ELSE}\ {\it expr}_2\ {\tt END}$.} 
\begin{tabbing}
$\forall(x).({\rm index}_{\it sel}(x) \Rightarrow ({\rm val}_{\it sel}({\it expr}, x) 
= {\rm val}_{\it sel}({\it expr}_{1}, x)$ 
\\ \quad $\iff {\rm val}_{\it sel}({\it expr'}, x) = \underline{{\tt TRUE}}))$.\\
$\forall(x).({\rm index}_{\it sel}(x) \Rightarrow ({\rm val}_{\it sel}({\it expr}, x) 
= {\rm val}_{\it sel}({\it expr}_{2}, x)$\\ 
\quad$\iff ({\rm val}_{\it sel}({\it expr'}, x) = \underline{{\tt FALSE}}
\vee {\rm val}_{\it sel}({\it expr'}, x) = \underline{{\tt NULL}})))$.
\end{tabbing}


\paragraph{{\bf Case:} ${\it expr} \coloneqq {\tt EXISTS}\ {\it subselect}$.} 

\begin{tabbing}
$\forall(x).({\rm index}_{\it sel}(x) \Rightarrow ({\rm val}_{\it sel}({\it expr}, x) = \underline{{\tt TRUE}}$\\
\quad $\iff 
\exists(y).({\rm index}_{\it subselect}(y))))$. \\
$\forall(x).({\rm index}_{\it sel}(x) \Rightarrow ({\rm val}_{\it sel}({\it expr}, x) = \underline{{\tt FALSE}}$\\ 
\quad $\iff 
\neg\exists(y).({\rm index}_{\it subselect}(y))))$. 
\end{tabbing}

\paragraph{{\bf Case:} ${\it expr} \coloneqq {\it subselect}$.}
Notice that a \emph{subselect} can only be used as an expression when it projects  one single item and  returns one single row. 
Then, in our methodology,  we first add the following 
\emph{proof goal} to our correctness proofs:
\begin{eqnarray*}
\lefteqn{{\rm o2f}_{\rm data}(\mathcal{D})}\\
& & \cup\  {\rm index}_{\rm def}(\mathcal{D})
\cup {\rm index}_{\rm def}({\it subselect}) \\
& & \cup\ \{\neg(\exists(x)({\rm index}_{\it subselect}(x) \\
& & \qquad \wedge \forall(y)(y \not= x \Rightarrow 
\neg({\rm index}_{\it subselect}(y)))))\}.
\end{eqnarray*}
\noindent Under the assumption
that this proof goal holds, 
the mapping ${\rm s2f}()$ generates the
following axioms: let ${\it expr'}$ be
the item projected by {\it subselect}, 
and let $w$ a new constant of the type of ${\it expr'}$.
Then, 

\begin{tabbing}
$\forall(x).({\rm index}_{\it sel}(x) \Rightarrow {\rm val}_{\it sel}({\it expr}, x) = w)$. \\
$\exists(x).({\rm index}_{\it subselect}(x) \land {\rm val}_{\it subselect}({\it expr'}, x) = w)$. 
\end{tabbing}

\end{document}